\documentclass[12pt]{article}
\usepackage{cite,epsfig,amssymb}

\def\R{R_{\odot} }
\def\O{{\cal O}}

\def\a{\alpha}
\def\b{\beta}
\def\g{\gamma}

\def\d{\delta}
\def\D{\Delta}
\def\e{\eta}
\def\l{\lambda}
\def\L{\Lambda}
\def\m{\mu}
\def\n{\nu}
\def\r{\rho}
\def\o{\omega}
\def\s{\sigma}
\def\S{\Sigma}

\def\p{\pi}
\def\ep{\epsilon}

\newcommand{\Ref}[1]{(\ref{#1})}

\def\ltsim{\lower3pt\hbox{$\, \buildrel < \over \sim \, $}}
\def\gtsim{\lower3pt\hbox{$\, \buildrel > \over \sim \, $}}
\def\be{\begin{equation}}
\def\ee{\end{equation}}
\def\ba{\begin{eqnarray}}
\def\ea{\end{eqnarray}}
\def\de{\partial}
\def\ga{\mathrel{\raise.3ex\hbox{$>$\kern-.75em\lower1ex\hbox{$\sim$}}}}
\def\la{\mathrel{\raise.3ex\hbox{$<$\kern-.75em\lower1ex\hbox{$\sim$}}}}

\begin{document}

\baselineskip=16pt   
\begin{titlepage}  
\rightline{IHES/P/02/90}
\rightline{OUTP/02/43P}
\rightline{hep-th/0212155}  
\rightline{December  2002}
\begin{center}  
  
\vspace{1cm}  
  
\Large {\bf Spherically symmetric spacetimes in massive gravity}  

\vspace*{5mm}  
\normalsize

{\bf Thibault Damour$^{a,}$\footnote{damour@ihes.fr}, Ian I. 
Kogan$^{b,}$\footnote{i.kogan@physics.ox.ac.uk} and  Antonios   
Papazoglou$^{c,}$\footnote{antpap@th.physik.uni-bonn.de}}

\medskip   
\medskip 
{\it $^a$Institut des Hautes \'Etudes Scientifiques}\\  
{\it 35 route de Chartres, 91440 Bures-sur-Yvette, France}  

{\it $^b$Theoretical Physics, Department of Physics, Oxford University}\\  
{\it 1 Keble Road, Oxford, OX1 3NP,  UK}  

{\it $^c$Physikalisches Institut der Universit\"at Bonn}\\  
{\it Nu\ss allee 12, D-53115 Bonn, Germany}  

\vskip0.5in \end{center}  
   
\centerline{\large\bf Abstract}  
 
We explore spherically symmetric stationary solutions, generated by ``stars''
with regular interiors, in purely massive gravity. We reexamine
the claim that the resummation of non-linear effects can cure, in a domain near the source,
 the discontinuity exhibited by the linearized theory as the mass $m$ of the
graviton tends to zero.  First, we find analytical
difficulties with this claim, which  appears not to be robust under slight changes
in the form of the mass term.  Second, by numerically exploring the inward
continuation  of the  class of asymptotically flat solutions, we find that, when $m$
is ``small'',  they  all end up in a singularity at a finite radius,
well outside the source,
instead of joining some conjectured ``continuous'' solution near the source.
We reopen, however, the possibility of reconciling massive gravity with phenomenology
by exhibiting a special class of solutions, with ``spontaneous symmetry breaking''
features, which are close, near the source, to general relativistic solutions and
asymptote,  for large radii,  a de Sitter solution of curvature $\sim  m^{2}$.

\vspace*{1mm}   

\end{titlepage}

\section{Introduction}

Over the past few years there  has been considerable discussion about theories with light 
massive
gravitons in their spectra. These kinds of theories first arose in the context of 
brane-world
models 
\cite{Kogan:1999wc,Gregory:2000jc,Kogan:2000cv,Kogan:2000xc,Kogan:2000vb,Kogan:2001yr,Dvali:2000hr,Dvali:2000xg}
and {\it at the linearized approximation} they predicted modifications of Newton's  
constant or even
of Newton's law itself  at cosmological scales. The evidence for a dark energy component 
in our universe and the associated cosmic acceleration 
\cite{Riess:1998cb,Perlmutter:1998np} made the study of these models rather  topical.

Indeed, generalizing these theories at the {\it non-linear level} 
\cite{Damour:2002ws}, revealed  that they naturally give rise to a
period of late time acceleration of the universe 
\cite{Deffayet:2000uy,Deffayet:2001pu,Damour:2002wu}. The non-linear analogue of a
collection of light massive gravitons is a theory with many interacting metrics, one of 
which only couples to the matter fields of our universe \cite{Damour:2002ws}. Depending on 
the particular form of the coupling of the several metrics,  the resulting
acceleration  could have interesting testable  differences  from a (scalar field) 
quintessence model, as for example anisotropic features \cite{Damour:2002wu}.

Despite the interest in these theories regarding cosmology, there are potentially 
dangerous issues associated
with the presence of extra polarization states of the massive gravitons 
\cite{vanDam:1970vg,Zakharov,Boulware:my}.
In particular, the massive gravitons have a scalar-like polarization state whose coupling
to matter does not depend on the mass of the graviton.
This scalar coupling to matter is formally analogous (in the linearized approximation) to 
a Jordan-Fierz-Brans-Dicke coupling
$\o = 0$. This coupling modifies (by a factor $ (2\o + 3)/(2 \o+4) = 3/4$)
the usual general relativistic relation between interaction of matter and light. If 
General Relativity (GR) is modified by a mass term,
or if GR is augmented by the addition of an extra massive graviton which
dominates\footnote{In the case where the massive graviton component were
sufficiently subdominant there would be no observational
discrepancy, but also no phenomenological interest for considering such a massive 
component.}
the matter couplings, the discrepancy between the theory of massive gravitons and GR in 
{\it e.g.} the bending of light by the Sun,
would be at the $25\%$ level, when the current observational accuracy is better that one 
part in $10^4$.

In addition to this (very serious) experimental difficulty, it has also been shown  that 
massive gravity has
serious theoretical defects \cite{Boulware:my,ACF,Deser}.  As soon as one goes beyond the linearized
level, massive gravity has
{\it six} degrees of freedom, instead of the expected five ($2 s + 1$, for a massive spin 
$s=2$).  This sixth degree of freedom is
problematic both because it represents a jump, with respect to the linearized theory, in 
the number of degrees of freedom,
and because its energy has no lower bound.
These arguments seem to exclude
the existence of dominantly contributing  massive gravitons, however small their mass 
might be. It  seems, in that case, that gravity is a unique
example of an ``isolated'' theory whose massive deformation can be excluded to infinite 
accuracy.

One could question several points in the arguments of
\cite{vanDam:1970vg,Zakharov,Boulware:my} recalled
above. The first one is that these arguments are based on the study of models containing 
an
{\it explicit} breaking of the (linearized) gauge invariance  $\delta 
h_{\mu\nu}=\de_{\mu}\xi_{\nu}+\de_{\nu}\xi_{\mu}$
by a  Pauli-Fierz mass term  for the light graviton(s), which reads:
\be
 \label{sumpf}
- { m^2 \over 4} (h_{\mu\nu}h^{\mu\nu}-h^2)
\ee
One could hope that the peculiar result of the discontinuity is linked to this explicit 
breaking and that  continuity in predictions
might be restored if the mass for the spin-2 field were generated spontaneously.
 For example, if  light gravitons are generated by the compactification of a higher 
dimensional theory, one could imagine a
 higher dimensional mechanism which would restore the continuity in the theory\footnote{For an alternative way of generating spontaneously masses for gravitons see \cite{Chamseddine:yu}.}.  We will 
not explore such possibilities
 in this paper, but study instead theories with explicit mass terms with the aim of
 clarifying whether such theories are indeed physically sick.

A second potentially weak point of the above argument is that the effects of
Pauli-Fierz mass terms were considered only around {\it flat space
backgrounds}. One could imagine that the discontinuity that was found
is a peculiarity of just the flat background and that if one considered
a background with curvature, some of the difficulties might be evaded. Indeed, it
was found that for instance in constant curvature backgrounds [$(A)dS$
spaces], the extra polarizations of the massive gravitons have a
coupling $\sim m/H$ where $m$ is the mass of the graviton and $H$ the ``Hubble'' constant 
of the $(A)dS$
space \cite{Higuchi:1986py,Higuchi:gz,Kogan:2000uy,Porrati:2000cp}. In
that case, the predictions of the massive theory were
indistinguishable\footnote{Note that at the quantum level the
discontinuity formally reappears \cite{Dilkes:2001av,Duff:2001zz}, but
as a quantum effect is suppressed and unobservable in any conceivable
near future experiment.} from the massless one as long as $m \ll
H$. This, however, prevents the massive gravitons from being cosmologically
interesting  since their Compton wavelength is much bigger than the
``Hubble'' radius of the ``bare'' cosmological constant.

In constant curvature backgrounds the parameter controlling  the smoothness of the limit is $m^2/{\cal R}$
where ${\cal R} \propto H^2$ is the curvature of the  (A)dS space.
One might conjecture
that the same is true  for any curved space and that the smoothness is
controlled by a parameter $\sim m^2/{\cal R}$, where ${\cal R}$  is some
curvature invariant. In the case of generic vacuum Einstein backgrounds
(with vanishing  scalar curvature and Ricci tensor)   ${\cal R}$
might be defined as some invariant of the Riemann tensor (such as the  square root of the
square of the  Riemann tensor).  In the case of a Schwarzschild background,
${\cal R} \sim R_S/ R^3$,  where $R_S$ is the Schwarzschild radius of the source. Then
a smooth limit might be expected to  exist  for distances $R \ll  R_p$, where the upper
bound $R_p$  of the expected interval of existence of smooth solutions is:
\be
R_p={R_S \over (mR_S)^{2/3}}  = ( {\l}_m^2    R_S )^{1/3}
\label{pscale}
\ee
where ${\l}_m \equiv m^{-1}$ denotes the Compton wavelength associated to the
mass $m$.

A similar conjecture was made a long time ago by Vainshtein \cite{Vainshtein:sx} based, however,  on
a different argument. That paper made two basic points:
(i) it questioned the use of  perturbation theory (and especially of the  linearized approximation)
 in the derivation of observable
consequences of massive gravity by showing that non-linear effects,   proportional to a
{\it negative} power of $m^2$, are important in a wide domain around the source,
 and (ii) it conjectured for the first time that  the resummation of non-linear effects  might actually
 restore continuity  in a domain near the source.  More precisely, this paper sketched the
 construction of solutions      which can be expressed, at least within some
 intermediate range of distances $R_S \ll \R  \leq R \ll R_V$, as a series in {\it positive} powers of
 $m^2$ and whose
leading term coincides with the one of GR.  The interval where this
expansion was constructed was bounded on the left by the radius of the star, $\R$,
 and on the right  by the length scale:
 \be
R_V={R_S \over (mR_S)^{4/5}}  =    ( {\l}_m^4    R_S )^{1/5}
\label{Vscale}
\ee
 which differs from the one obtained by
the  conjecture made in the previous paragraph.

 In the simple example of a massive graviton with Compton wavelength of the order of the universe's horizon
and of a source as massive as the Sun,  the  distance scale \Ref{Vscale} is much bigger than the
distance scales on which relativistic predictions of GR are tested. Thus, if the conjecture
were correct, the massive graviton proposal would not be excluded.
  [Actually, the same holds if continuity is restored on the scale \Ref{pscale}, though $R_p \ll R_V$.]
The arguments of \cite{Vainshtein:sx} were repeated, with more intermediate details,
in the recent paper of \cite{Deffayet:2001uk}. This paper also provided
   a  cosmological analogue of the
continuity-discontinuity interplay  in the context of the brane-induced
gravity models (see also \cite{Lue:2001gc} for a cosmic string example for the same kind of
brane models).

 However, the arguments of    \cite{Vainshtein:sx,Deffayet:2001uk} are questionable,
 and one of the main thrusts of the present paper will be to show that they actually are inconclusive.
The questionable aspects concern both the {\it local existence} of  continuous
solutions, their {\it global relevance}, and even their  {\it global existence}.
For instance, we shall show that the action principle postulated in   \cite{Deffayet:2001uk}
  {\it does not admit },  in the intermediate    range  $R_S \ll \R  \leq R \ll R_V$,
local solutions of the type conjectured in \cite{Vainshtein:sx} for restoring continuity.
 Moreover, it is not at all clear if  such {\it local} solutions, when they do exist (for
 some other action principles), can be {\it globally} continued to
 the asymptotically flat ones of the
linearized theory, a concern first raised in \cite{Boulware:my} and confirmed later in \cite{Jun:hg}. Actually, it is not
even clear  if these solutions are indeed solutions, {\it i.e.} if they do not contain
naked singularities at some finite distance. 

Let us note here that there have been several papers  \cite{Gruzinov:2001hp,Porrati:2002cp,Middleton:2002qa,Dvali:2002vf,Lue:2002sw} where the Vainshtein idea was investigated in brane-induced gravity for static spherically symmetric sources and novel phenomenology was proposed \cite{Dvali:2002vf,Lue:2002sw} \footnote{See \cite{Giannakis:2001jg} for a demonstration of the discontinuity at the linearized level for spherically symmetric sources in brane-induced gravity models. For the same models, see also \cite{Kofinas:2001qd,Kofinas:2002gq} for special spherically symmetric solutions.}. In our view, the approximations made in these papers, although plausible, do not address the global properties of the solutions.  We will not explore the specific brane-induced gravity model in this paper.

In the present paper we will discuss, within the context of purely massive gravity,
 the claims of \cite{Vainshtein:sx} as well
as the more expanded version of them that appeared in \cite{Deffayet:2001uk}.
We will analyse in detail the procedure that was used in
\cite{Vainshtein:sx} to obtain the results, shedding light to some obscure points, and
repeating the same calculation for several mass terms. We will show that the method
has serious limitations and that there are contradictions between the statements made
in  \cite{Vainshtein:sx} and in \cite{Deffayet:2001uk}.  Indeed, we shall point out that
the field equations studied in  \cite{Vainshtein:sx} do not derive from the action
principle postulated in  \cite{Deffayet:2001uk}. Actually, they cannot be derived from {\it any}
action principle. We shall show that, if one starts from the simple action principle written
in  \cite{Deffayet:2001uk}, there exist {\it no} expansions of the type postulated
in \cite{Vainshtein:sx} as a way to restore continuity. However, we shall show that
such expansions do exist for other action principles (and do exist for the field equations
written down in \cite{Vainshtein:sx}, after correction of some misprints).
We will also see that the scale $R_V$ up to which the expansion of
\cite{Vainshtein:sx} makes sense, is not universal for different potential terms.
We will then study (numerically) the {\it global aspects} of such solutions.  Our numerical simulations
strongly suggest that: (i) the Vainshtein-type local (approximate)  solutions  {\it do not match},
as they were supposed to,  the asymptotically flat approximate solutions generated by
normal perturbation theory (in agreement with \cite{Jun:hg}), and in fact that (ii) the asymptotically flat solutions {\it cannot be
extended} inwards to globally regular solutions. Even if we forget about the issue of
mass-continuity, we find the striking result that all   asymptotically flat solutions
run into naked singularities  as the radial coordinate $R$ decreases.

However, our conclusions will not all be negative.  Indeed, we shall explicitly construct
some globally regular solutions of massive gravity which are continuously connected,
when $m \to 0$, to GR solutions, and which are phenomenologically  consistent
with experimental tests of relativistic gravity. The global solutions we shall (numerically) construct
are  generalizations (to the inclusion of a central source) of the black-hole-type
solutions of massive gravity constructed long ago by Salam and Strathdee  \cite{Salam:1976as}.
[The latter solutions were generalized by Isham and Storey \cite{Isham:1977rj} for a
particular class of bigravity theories  \cite{Isham:gm}.]
The main difference between these solutions and the ones studied in Refs.
\cite{vanDam:1970vg,Zakharov,Boulware:my,Vainshtein:sx}  is the behaviour
at infinity.  The solutions are not required to be asymptotically flat, but instead to
match to a cosmological solution of massive gravity (which is de Sitter, in the
case at hand).  In fact, a general argument of
\cite{Damour:2002ws} (see section 4 there) has shown how to construct (at least
for a limited time) general classes of  solutions  which are continuously connected,
 when $m \to 0$, to GR solutions representing local gravitating systems (such as the
 solar system), embedded in some global cosmological background.
 Another big difference with the
 solutions studied in   \cite{vanDam:1970vg,Zakharov,Boulware:my,Vainshtein:sx},
 and in the first part of this paper, is that these cosmologically-matched solutions
 are of a special ``symmetry-breaking'' type (see below).

 As we shall discuss in our conclusions, our results leave open several important issues
 which must be tackled before a firm conclusion can be reached concerning the
 physical consistency (or inconsistency) of massive gravity theories.

\section{Action and potentials for massive gravity}
\label{anp}

Our starting point is a generic action for (four dimensional) massive gravity. This can
be obtained, as it was shown in \cite{Salam:1976as,Damour:2002ws}, from a four dimensional
bigravity action if we send the gravitational constant associated with  the second  metric
(the one not coupled to ``our world'')
to zero. Then the corresponding metric gets (formally) frozen, {\it i.e.} it becomes a non-dynamical Einstein space background.
We assume that this non-dynamical metric is {\it flat}, and we accordingly denote it by $ \bf f$. 
[We do not necessarily assume that the metric   ${\bf f}$      is written in Lorentzian coordinates;
{\it i.e.} the components $f_{\m\n}(x)$ are not necessarily assumed to be simply $ {\e}_{\m\n}$.]
The only remaining dynamical metric is  denoted ${\bf g}$. Then the effective action for the
purely massive gravity is: 
\be
\label{action}
{\cal S}={1 \over  16 \pi G}\int d^4x \left(\sqrt{-g} R[{\bf g}]-{m^2 \over 4} {\cal
V}({\bf f}^{-1}{\bf g})\right)+{\cal S}_{{\rm matt}}[{\bf g}]
\ee
where $G$ is the ``massive'' version of  Newton's constant
and $m$ a mass parameter, which for the specific
normalization of the potentials ${\cal V}$ which we will consider, is the canonically 
normalized Pauli-Fierz mass of the graviton. In the following, we explore the following 
possible potential terms:
\ba
&&{\cal V}^{(1)}=\sqrt{-g}\left\{{\rm tr}[(({\bf g}-{\bf f}){\bf f}^{-1})^2]-({\rm tr}[({\bf 
g}-{\bf f}){\bf f}^{-1}])^2 \right\}\label{V1}\\
&&{\cal V}^{(2)}=\sqrt{-f}\left\{{\rm tr}[(({\bf g}-{\bf f}){\bf f}^{-1})^2]-({\rm 
tr}[({\bf g}-{\bf f}){\bf f}^{-1}])^2 \right\}\label{V2}\\
&&{\cal V}^{(3)}=\sqrt{-g}\left\{{\rm tr}[(({\bf g}^{-1}-{\bf f}^{-1}){\bf f})^2]-({\rm 
tr}[({\bf g}^{-1}-{\bf f}^{-1}){\bf f}])^2 \right\}\label{V3}\\
&&{\cal V}^{(4)}=\sqrt{-f}\left\{{\rm tr}[(({\bf g}^{-1}-{\bf f}^{-1}){\bf f})^2]-({\rm 
tr}[({\bf g}^{-1}-{\bf f}^{-1}){\bf f}])^2 \right\}\label{V4}
\ea
Note that the above quantities are not scalars, but scalar densities. Comparing with the 
notation of \cite{Damour:2002ws}, ${\cal V}=({\rm density})V$ where $V$ is a scalar. All 
of the four above potentials are of the Pauli-Fierz type near ${\bf g} \approx {\bf f}$. Note
that the first one was written to be the starting point in \cite{Deffayet:2001uk}.

Each contribution ${\cal S}_x$ to the action  gives rise to a corresponding   contribution
to the energy-momentum tensor:  $T^x_{\m\n} \equiv -( 2/\sqrt{-g})~ \delta {\cal S}_x / \delta g^{\m\n}$.
The ``gravitational'' energy-momentum tensors  arising from the mass terms will be
denoted  $T^{(g)}_{\m\n}$. For the specific mass terms above, they are respectively:
\ba
&&T_{\mu \nu}^{(g1)}={m^2 \over 16\pi G}\left[-{1 \over 4} g_{\mu
\nu}(h^{\alpha\beta}h_{\alpha \beta}-h^2)-(g_{\mu \kappa}g_{\nu \lambda}h^{\kappa 
\lambda}-hg_{\mu \kappa}g_{\nu \lambda}f^{\kappa \lambda})\right]\\
&&T_{\mu \nu}^{(g2)}=-{m^2\over 16\pi G}{\sqrt{-f} \over \sqrt{-g}}(g_{\mu \kappa}
g_{\nu \lambda}h^{\kappa \lambda}-h g_{\mu \kappa}g_{\nu \lambda}f^{\kappa \lambda})\\
&&T_{\mu \nu}^{(g3)}={m^2\over 16\pi G}\left[-{1 \over 4} g_{\mu
\nu}(H^{\alpha\beta}H_{\alpha \beta}-H^2)+(H_{\mu \nu}-H f_{\mu \nu})\right]\\
&&T_{\mu \nu}^{(g4)}={m^2\over 16\pi G}{\sqrt{-f} \over \sqrt{-g}}(H_{\mu \nu}-H
f_{\mu \nu})
\ea
where we have defined $h_{\mu \nu}=g_{\mu \nu}-f_{\mu \nu}$ and $H^{\mu \nu}=g^{\mu
\nu}-f^{\mu \nu}$. The indices of $h_{\mu \nu}$ are raised by $f^{\mu \nu}$ and the 
ones of  $H^{\mu \nu}$ are lowered by $f_{\mu \nu}$. Note that $H^{\mu \nu}=-h^{\mu 
\nu}+h^{\mu \kappa}h^{\nu}_{\kappa}+\cdots$, which explains the sign differences in the 
above expressions.

In this paper we wish to consider spherically symmetric stationary (SSS) solutions for these mass terms, {\it i.e.}
SSS solutions of:
\be
\label{graveqs}
R_{\m\n} - {1 \over 2} R g_{\m\n} = 8 \p G ( T_{\m\n}^{(g)} + T_{\m\n}^{(\rm matt)})
\ee
Note that the separate diffeomorphism invariance of the matter action    $ {\cal S}_{{\rm matt}} $
(which is assumed to couple only to ${\bf g}$)
implies the separate conservation  (on matter shell) of the material energy-momentum tensor:
\be
\label{matcons}
\nabla^{\mu} T^{\rm (matt)}_{\mu\nu}=  0
\ee
A consequence of Eqs. \Ref{graveqs}, \Ref{matcons} and of the Bianchi identities
is then the separate conservation  of the gravitational energy tensor:
\be
\label{gravcons}
  \nabla^{\mu} T^{(g)}_{\mu\nu}=  0
\ee

\section{Poincar\'e-covariant perturbation theory}
\label{poincarepert}

Note that all the actions \Ref{action} admit as exact solution, in the absence of matter,  the
``trivial vacuum'' $ {\bf g} = {\bf f}$. [See, however, section 8 below for a discussion
of non-trivial vacua.] When representing the non-dynamical flat
background metric      ${\bf f}$  in Lorentzian coordinates, {\it i.e.} $f_{\m\n} = {\e}_{\m\n}$,
one can then develop a   Poincar\'e-covariant  perturbation theory:
$  h_{\mu \nu}=g_{\mu \nu}-{\e}_{\mu \nu} =   h_{\mu \nu}^{(1)} +  h_{\mu \nu}^{(2)} + \cdots$.
Within such an approach it is implicitly required that the massive gravitational
field      $  h_{\mu \nu}$ decay at infinity, so that:
\be
\label{bc}
  g_{\mu \nu} \to {\e}_{\mu \nu}~~ , ~~ {\rm as} ~~  r \to \infty
\ee
In the first part of this paper, we shall impose this standard requirement, the problem
being to assess the existence and continuity of solutions of massive gravity matching
the trivial vacuum at infinity. [However, we shall relax this requirement in the second
part of the paper.]

Before entering the details of our investigation of  non-linear SSS solutions, let us,
as a warmup, recall the basic features of perturbation theory. At the linearized approximation, the equations of motion read:
\be
-\Box h_{\mu \nu} + h_{\mu ,\lambda \nu}^{\lambda}+ h_{\nu ,\lambda \mu}^{\lambda}
-\eta_{\mu \nu}h^{\kappa \lambda}_{\phantom{\kappa \lambda},\kappa \lambda}-h_{,\mu \nu}
+\eta_{\mu \nu}\Box h+ m^2 (h_{\mu \nu}-\alpha \eta_{\mu \nu} h)=16 \pi G T_{\mu \nu} \label{BD}
\ee
Here all index raisings and lowerings are made with $\eta_{\mu \nu}$, and the source term
$T_{\mu \nu}$ is equal, at the linearized approximation, to the matter energy-momentum tensor. [See below, 
for the generalization of Eq. \Ref{BD} to the non-trivial vacuum cases.]
Following \cite{Boulware:my} we have introduced  a
parameter $\a$ to study the special r\^ole of the Pauli-Fierz mass term, which corresponds
to ${\a}_{\rm PF} = 1$.

Let us  recall  the consequences of Eq. \ref{BD}). The divergence of this equation yields, when using   the
conservation of the source, ${\de}^{\n} T_{\m\n} = 0$, the constraint:
\be
\label{BDdiv}
 h_{\mu \nu}^{\phantom{\mu \nu},\nu}=\alpha h_{,\mu}
 \ee
Then the trace gives:
\be
2(1-\alpha)\Box h  +m^2 (1-4\a)h= 16 \pi G T \label{BDtrace}
\ee
From the latter equation we see that if $\alpha=1$, {\it i.e.} when we sit on the Pauli-Fierz point, the trace of
$h_{\mu \nu}$ is locally determined by a constraint:
\be
h=-{16 \pi G T \over 3 m^2 }
\ee
and thus substituting in Eq. (\ref{BD}) one gets:
\be
\label{BDresult}
{1 \over 16 \pi G }h_{\mu \nu}= {1 \over -\Box +m^2}\left(T_{\mu \nu}-{1 \over 3}T \eta_{\mu \nu}\right)
+{1 \over 3m^2}~{1 \over -\Box +m^2}T_{,\mu \nu}
\ee
This result for the massive gravity field in terms of the source exhibits the difficulties of massive gravity.
First, there are the phenomenological difficulties associated with the $1/3$ instead of the Einsteinian $1/2$ factor
in front of $ T \eta_{\mu \nu}$.
Second,   there are the theoretical difficulties associated with the presence of a factor $m^{-2}$
in some terms of the solution.

Before discussing further these difficulties beyond the linearized level, let us also briefly recall the
results of     \cite{Boulware:my,Deser}    concerning the  non-Pauli-Fierz mass terms, {\it i.e.} the case
 $\alpha \neq 1$. In this case,
 we see from  Eq. (\ref{BDtrace}) that the trace of $h_{\mu \nu}$ becomes a dynamical field, {\it i.e.}  a new
 degree of freedom. Then the corresponding result for the expression of $h_{\mu \nu}$ in terms of the source is:
\ba
\label{ghost}
{1 \over 16 \pi G }h_{\mu \nu}= {1 \over -\Box +m^2}\left(T_{\mu \nu}-{1 \over 3}T \eta_{\mu \nu} \right)
-{1 \over 6}\eta_{\mu \nu}{1 \over -\Box +m_0^2}T\nonumber \\
+ {2 \alpha-1 \over 2(1-\alpha)}{1 \over -\Box +m^2}{1 \over -\Box +m_0^2}T_{,\mu \nu}
\ea
where
\be
m_0^2=m^2 {4\alpha -1 \over 2(1-\alpha)}
\ee
 is the mass of the extra scalar degree of freedom.
[It is non-tachyonic if $\a$ is in the interval $ 1/4 \leq \a < 1$]. Note
that this field is always {\it ghostlike} (independently of the value of $\a$),
 as seen by considering the various contributions to the
action, $h_{\mu \nu} T^{\mu \nu}$, where the  term $-(1/6) \eta_{\mu \nu} ( -\Box +m_0^2)^{-1}T$
contributes with the opposite sign from the first term (note that the last, double gradient term,
does not contribute, after integration, because of the conservation of   the source, ${\de}^{\n} T_{\m\n} = 0$).
As emphasized in   \cite{Boulware:my,Deser}   the
case    $\alpha \neq 1$   does not exhibit the difficulties of the Pauli-Fierz case.
On the one hand, the expression \Ref{ghost} contains no dangerous denominators, vanishing with $m^2$,
and on the other hand, it smoothly merges (thanks to the identity $1/3 + 1/6 \equiv 1/2$)
 into the GR result for the integral of   $h_{\mu \nu} T^{\mu \nu}$
when     $m^2 \to 0$.  Those nice continuity properties of the case    $\alpha \neq 1$ pointed out in
 \cite{Boulware:my,Deser}   are particularly
evident in the case $\a =1/2$, but hold in all cases    $\alpha \neq 1$ ,  independently of
whether the extra scalar degree of freedom is tachyonic or not.  [These continuity properties have
been recently further studied in    \cite{Babak:2002uz,Sami:2002qg}].
 Nevertheless, the presence
 of a new ghostlike degree of freedom makes the theory for $\alpha \neq 1$ pathological as a
 quantum theory. In view of this quantum ghost instability, we restrict ourselves to the
 ghost-free Pauli-Fierz case.  We note in passing that Pauli-Fierz mass terms naturally
 arise from higher-dimensional gravity models. Indeed, the structure of the Pauli-Fierz
 mass term is already encoded in the structure of the gradient terms in Einstein's action (see, {\it e.g.},
 section 3.1 of    \cite{Damour:2002ws} ).

 Let us now briefly indicate how the Poincar\'e-covariant perturbation theory would proceed
 beyond the linear approximation.   As is done in  the Poincar\'e-covariant perturbation
 theory in GR one can define an effective energy tensor $T^{\rm eff}_{\m\n}$, which
 combines the matter energy tensor with the non-linear parts $\sim \de \de h h + m^2 h h + \cdots$
  of the left-hand side
 of the exact field equations  \Ref{graveqs}, such that the full field equations read
 as the linearized ones, Eq. \Ref{BD}, with the replacement $ T_{\m\n} \to   T^{\rm eff}_{\m\n}$.
 This effective energy tensor is not exactly conserved but rather satisfies a relation
 of the type     ${\de}^{\n} T^{\rm eff}_{\m\n} = m^2 ( \de h h + \cdots)_{\m}$.
 This modifies both Eq.   \Ref{BDdiv} and Eq.     \Ref{BDtrace}, and thereby
 Eq.   \Ref{BDresult},  by non-linear terms.  By iteration, one can then deduce the
 following structure for the non-linearity expansion of the solution:
 \be
 \label{expansion}
 h \sim \left(1 + { 1 \over mR}  + { 1 \over (mR)^2}\right) U + \left(1 + { 1 \over mR} + \cdots + { 1 \over (mR)^6}\right) U^2
 + \cdots
 \ee
Here, we have denoted by $U$ a typical gravitational potential  which is, outside the source
and for distances $R \ll {\l}_m $,
of order  $U \sim     R_S/R$.   The expansion \Ref{expansion} contains increasingly high inverse
powers of $m$. We can rewrite it as a sum of terms of the type $ ( R_{p,q}/R)^n $ with a sequence
of length scales defined as:
\be
\label{pqscale}
R_{p,q} \equiv      ( {\l}_m^p    R_S^q )^{1/(p+q)}
\ee
where, as above, $ {\l}_m \equiv m^{-1}$.

The expansion above can numerically make sense only  if the radius $R$ is (much) {\it  larger} than
all the relevant length scales    $  R_{p,q}$. In view of the extremely large difference between
 $ {\l}_m$ ($\sim 10^{28}$ cm, say) and $ R_S$ ($\sim 10^5$ cm, say, for the Sun) the various
 length scales tend to be extremely large, which confirms the first point made in   \cite {Vainshtein:sx},
 namely the irrelevance of perturbation theory for describing the massive gravitational
 field near the source.  For instance,  the light deflection by the Sun probes the gravitational field
 just outside the radius of the Sun, while the expansion above represents $h$ there as a sum of terms, starting
 with a small  first term $ U \sim R_S/R \sim 10^{-6}$, but continuing with extremely large
  ``corrections''  such as $U^2/( mR)^2 \sim 10^{+22}$ !

 Actually, the straightforward perturbation expansion \Ref{expansion} can be significantly improved
 by gauging away some of the worse terms. Indeed, the primitive source for the ``bad'' inverse
 powers of $m^2$ is the last term in Eq. \Ref{BDresult}. However, this term is mainly a ``gauge term'',
 in the sense that a suitable coordinate transformation $ x_{\rm old}^{\m} \to x_{\rm new}^{\m} $
  can
 remove this term from the physical  ``massive" metric $g_{\m\n}(x)$, at the cost of introducing it
 in the non-dynamical flat metric $f_{\m\n}(x)$ (now written in the ``curved'' coordinate system
  $ x_{\rm new}^{\m}$  instead of the original Lorentzian coordinate system).
  However, the crucial point is that this ``gauging away'' cannot remove all inverse powers of
  $m^2$.   Indeed, if we write $   x_{\rm old}^{\m} =  x_{\rm new}^{\m} +   {\xi}^{\m} (  x_{\rm new})$,
  we have   $ g^{\rm new}_{\m\n}(x) = {\de}_{\m}( x^{\a} +   {\xi}^{\a})   {\de}_{\n}( x^{\b} +   {\xi}^{\b})
  g^{\rm old}_{\a\b}(x + \xi)$, so that the expansion of the new field $ h^{\rm new} \equiv g ^{\rm new}
  -\e$ contains, besides the terms $ \de \xi$ which can remove the last term in  Eq.   \Ref{BDresult},
 the new terms $ \sim   \de \xi  h + \de \xi   \de \xi$. Therefore, with $\xi \sim \de U/m^2
 \sim U/(m^2 R)$ chosen
 to remove the worst term $\propto m^{-2}$ in  Eq.   \Ref{BDresult} , we still end up with several
 terms that contain inverse powers of   $m^2$, namely:
 \be
 \label{improvedexpansion}
 h^{\rm new} \sim U +     \left(1 + { 1 \over mR}+ \cdots + { 1 \over (mR)^4}\right)   U^2    + \cdots
 \ee
 One could continue to apply suitable gauge transformations to remove some of the worse
 terms arising from computing higher iterations of $h$, but it is clear that once some
 negative powers of $m^2$ have entered $h$ they will keep generating higher powers
 in higher iterations. In view of the vast difference in scale between      $ {\l}_m$ ( $\sim 10^{28}$ cm)
 and phenomenologically tested scales $R$ ($\sim 10^{11}$ cm for the Sun radius) the dangerous
inverse denominators are of order of powers of $   (mR)^{-2 } \sim 10^{+34}$   and completely outgrow the
small numerators which are powers of $U \sim 10^{-6}$.   Finally, the improved expansion
\Ref{improvedexpansion}   is again irrelevant for describing the massive gravitational
field near the source, and {\it can only be valid in some neighbourhood of infinity}.
On the other hand, note that if we consider, for simplicity, the case of
stationary sources the basic ``Newtonian'' potential $U$ in the expansions above
will always be a Yukawa-type potential (obtained by the action of $(-\Box + m^2)^{-1}$
on the material source $T$). This potential will decrease exponentially as $R \to \infty$.
Therefore it is clear that perturbation theory will be formally well defined, to all orders, when $ R \to \infty$,
and will generate a numerically meaningful series (say in the whole domain $ R > m^{-1}$).
This series can be thought of as uniquely defining a specific solution of massive gravity.
The main question of concern being to know whether this solution, initially defined only in
a neighbourhood of infinity, can be continued into a regular solution everywhere.

Before leaving the topic of Poincar\'e-covariant  perturbation theory, it is important to notice
the following. If we apply this perturbation theory to the particular case of {\it static} spherically
symmetric  sources (with no motion in the Lorentzian coordinate system),
 it is clear that each term of perturbation theory will be static and spherically symmetric.
 In particular, each term will be invariant under time-reversal $ t \to -t$. This implies that,
 in Lorentzian coordinates ({\it i.e.} with $f_{\m\n} = {\e}_{\m\n}$,) there cannot be
 {\it off-diagonal} terms $ g_{0i}$ in the massive metric. Therefore, the exact
 solution defined (as above) by perturbation theory can be written in a {\it bi-diagonal}
 form, say:
 \be
 \label{abc}
 {\bf f } = -dt^2 + dr^2 + r^2   d\Omega^2 ~~~ ,~~~
  {\bf g} = - a^2 dt^2 + b^2 dr^2 + c^2   d\Omega^2 ,
 \ee
 with $a,b,c$ some functions
 of $r$, and $  d\Omega^2 = d \theta^2 + \sin^2 \theta d\phi^2$.
  We shall later reconsider this bi-diagonality.

\section{The $\lambda$, $\mu$, $\nu$ gauge}

  When discussing SSS  solutions,
it is instructive to work in two different gauges for the metrics.
We have just introduced  the  $a,b,c$ gauge \Ref{abc} (which will be
used below to introduce some convenient variables $c$, $\bar{c}$, $\bar{b}$).
It is also convenient to work in the $\lambda$, $\mu$,
$\nu$ gauge introduced now.
The latter is useful for comparing our results  with the literature,  while the former is similar
to the gauges used in cosmological studies, and has also the merit to give a convenient
 Lagrangian formulation to the system.
Let us write the observable line element  $ds^2=g_{\mu \nu}dx^{\mu}dx^{\nu}$ and the
reference flat metric one  $ds_{\rm fl}^2=f_{\mu \nu}dx^{\mu}dx^{\nu}$ in the following 
 (``Schwarzschild'') gauge:
\ba
&&ds^2=-e^{\nu(R)}dt^2+e^{\lambda(R)}dR^2+R^2 d\Omega^2\label{la1}\\
&&ds_{\rm fl}^2=-dt^2 +\left(1-{R\mu' \over 2}\right)^2e^{-\mu(R)}dR^2+e^{-\mu(R)}R^2 
d\Omega^2\label{la2}
\ea
where $'\equiv {d \over dR}$.
This gauge was used  in part of the literature of spherically
symmetric solutions in massive gravity.
It has the advantage of separating the directly observable gravitational variables
$\n(R),~ \l(R)$ from the ``gauge'' function $\m(R)$. The coordinate redefinition $r=R e^{-\mu/2}$ shows
trivially that the second metric is indeed flat, and that the form \Ref{la1},   \Ref{la2} is
actually equivalent to the  form \Ref{abc}.  [See below for the explicit link between the
two sets of variables.]

 The most general matter energy-momentum tensor which respects the spherical symmetry is 
of the form:
\be
T_{{\rm (matt)}~\nu}^{\phantom{{\rm (matt)}}~\mu}={\rm diag}(-\rho,P_r,P_t,P_t)
\ee
where we have left open the possibility of having different different radial $P_r$  and tangential $P_t$ pressures.
In the case of a ``fluid'' source, we shall have  $P_r = P_t$. Then the $(t,t)$
and $(R,R)$ components of the  Einstein equations read:
\ba
&&e^{\nu-\lambda}\left[{\lambda' \over R}+{1 \over R^2}(e^{\lambda}-1)\right]=m^2 
f_t(\lambda,\mu,\nu,\mu',R)+8 \pi G \rho ~e^{\nu}\label{laeq}\\
&&{\nu' \over R}+{1 \over R^2}(1-e^{\lambda})=m^2 f_{R}(\lambda,\mu,\nu,\mu',R)+8 \pi G 
P_r ~e^{\lambda}\label{nueq}
\ea
where the quantities $f_t$ and $f_R$ are proportional to the $(t,t)$ and $(R,R)$ 
components of the energy momentum tensor $T^{(g)}_{\mu\nu}$ generated by the potential 
term, and are defined as:
\be
f_t \equiv {1 \over 4}{1 \over  \sqrt{-g}}{\delta {\cal V} \over \delta g^{tt}} \equiv  {8 \pi G
\over m^2}T^{(g)}_{tt} ~~~ {\rm and}~~~ f_R \equiv {1 \over 4} {1 \over \sqrt{-g}}{\delta {\cal
V} \over \delta g^{RR}}\equiv  {8 \pi G \over m^2}T^{(g)}_{RR}
\ee

From the conservation of the energy-momentum tensor generated by the mass term, {\it i.e.} 
from $\nabla^{\mu} T^{(g)}_{\mu\kappa}$=0, we obtain a third equation from the $\kappa=r$
component (the $\kappa=t,\theta,\phi$ components are identically zero). Defining:
\be
f_g \equiv {8 \pi G \over m^2}\nabla^{\mu} T^{(g)}_{\mu r}
\ee
we have the following constraint equation for $m \neq 0$:
\be
f_g(\lambda,\mu,\nu,\lambda',\mu',\nu',\mu'',R)=0\label{coneq}
\ee
Note that this equation has no $m$ dependence. In the Appendix A we present the functions 
$f_t$, $f_R$ and $f_g$ for all the potentials that we are considering.

Finally, from the conservation of the matter energy-momentum tensor we have the final 
equation which closes the system:
\be
\label{conseqaniso}
2 P_r'+{1 \over R}(3P_r-\rho-4P_t)+\left(\nu'+{1 \over R}\right)(\rho+P_r)=0
\ee

Note that in the ``fluid'' case, {\it i.e.} for an isotropic pressure, the latter equation
simplifies to:
\be
P' = - {{\n}' \over 2} ( \rho+P)
\ee

\subsection{The General Relativity $m=0$ limit}

In the GR limit, one must discard Eq. \Ref{coneq}, and set  $m^2$ to zero  in
Einstein's equations     \Ref{laeq}, \Ref{nueq}.
It will be helpful to remind ourselves of the solution of the Einstein equations in GR for 
a star of constant density $\rho=\rho_0$ and isotropic pressure $P=P_r=P_t$. In this case
(considered here for simplicity),
the exterior (vacuum) solution , after absorbing a constant by a time rescaling, reads:
\be
e^{\nu}=1-{R_S \over R}~~~~,~~~~\lambda=-\nu
\ee
where $R_S \equiv 2GM$ is the Schwarzschild radius and appears here as an integration 
constant.

On the other hand, inside the star, eq. (\ref{laeq}) can be integrated to:
\be
e^{-\lambda}=1-{2G m(R) \over R} \label{GRintla}
\ee
where   we have defined $m(R)={4 \over 3}\pi \rho_0 R^3$. Note that we have
imposed the constraint $\l (R=0) = 0$, which is necessary to avoid
a   conical singularity at $R \to 0$.         If we denote the radius of the star by
$ \R$, matching to the external solution gives the links
 $m(R_{\odot})=M$ and    $R_S={8 \over 3}\pi G \rho_0 R_{\odot}^3$.
  We then can solve the remaining equations and obtain the well known
result for the pressure:
\be
P=\rho_0 \frac{\sqrt{1-{R_S \over R_{\odot}^3}R^2}-\sqrt{1-{R_S \over 
R_{\odot}}}}{3\sqrt{1-{R_S \over R_{\odot}}}-\sqrt{1-{R_S \over R_{\odot}^3}R^2}}
\ee
and the redshift function $\nu$ (corresponding to the {\it same} rescaled time as the external solution
above):
\be
e^{\nu}=\left({3 \over 2}\sqrt{1-{R_S \over R_{\odot}}}-{1 \over 2}\sqrt{1-{R_S \over 
R_{\odot}^3}R^2}\right)^2 \label{GRintnu}
\ee

\subsection{The perturbative limit for $m \neq 0$}
\label{pertsol}

Let us now see how the results change if we consider the massive theory. We will firstly 
deal with the linearized equations of motion with the aim of
understanding analytically how one can match the {\it unique} decaying exterior
solution into a regular interior one. In other words we will assume that
$\lambda,\mu,\nu \ll 1$. We will solve these linearized equations in the interior and the
exterior of the star and then match them to determine completely the various integration 
constants.

\subsubsection{The exterior star solution}

The equations (\ref{laeq}), (\ref{nueq}), (\ref{coneq}) at linear order in $\lambda,\mu,\nu$
 are the following  (for all four mass terms):
\ba
&&{\lambda' \over R}+{\lambda \over R^2}=-{m^2 \over 2}(\lambda+3\mu+R \mu')\\
&&{\nu' \over R}-{\lambda \over R^2}={m^2 \over 2}(\nu+2\mu)\\
&&\nu'={2\lambda \over R}
\ea
Note that the linearization of the equations has removed entirely the function $\mu$ from 
the constraint equation (\ref{coneq}). These equations, give a simple second-order differential
equation for $\nu$:
\be
\nu''+{2\over R}\nu'-m^2 \nu=0 \label{extnu}
\ee
Then the other variables are obtained from $\n$ via:
\be
\label{lmintermsofn}
\l = {R \over 2} {\n}' ~~~ , ~~~
\m = {\nu' \over 2 m^2 R} - {\n \over 2}
\ee
The generic solution of  the $\n$ equation above contains two integration constants, $C_1$
and $C_1'$.  The solution proportional to  $C_1$ is exponentially decaying at infinity, while the
one proportional to $C_1'$ is exponentially growing.  In keeping with our boundary condition  \Ref{bc}
we shall reject the exponentially growing solution and consider only the other one,
namely:
\ba
&&\nu=-{C_1 \over R}e^{-mR}\label{nupert}\\
&&\lambda={m C_1 \over 2}\left(1+{1 \over mR}\right)e^{-mR}\label{lapert}\\
&&\mu={C_1 \over 2R}\left(1+{1 \over mR}+{1 \over (mR)^2}\right)e^{-mR}\label{mupert}
\ea
Note that, if needed, the exponentially growing solution is simply obtained from
the latter solution by changing $ m \to -m$.
The integration constant   $C_1$ is proportional to the Schwarzschild radius    $R_S \equiv 2 G M$
of the source (the precise link will be seen later when we match with
the interior).  In the region  $R \ll m^{-1}$ we
obtain the limits of the above solution for $\l,\m,\n$:
\be
\label{disc}
\nu \approx -{C_1 \over R}~~~,~~~\lambda \approx -{ \n \over 2}
 \approx {C_1 \over 2R}~~~,~~~\mu \approx {C_1
\over 2m^2 R^3}
\ee
Note that the second relation differs from the corresponding GR relation
$\l \approx - \n$.
These show the well known fact that there is a finite discontinuity between this metric 
and the one obtained in GR, independent of the mass of the graviton as was shown in 
\cite{Boulware:my}. Additionally, the last relation shows that, in the $\l,\m,\n$ gauge, and
at the linearized approximation, there is a  dangerous denominator  $m^2$
only in the ``gauge'' function $\m$.
Actually, one can see that the gauge function    $\m$  is similar to the gradient   $\de \xi$
of the  coordinate transformation that we introduced above to remove the leading
${\cal O} (m^{-2})$ terms in the (linearized) perturbative field $h$.  We had above
$\de \xi \sim U/(mR)^2$ which indeed corresponds to $ \m \sim R_S/(m^2 R^3)$.
Therefore, as above, the choice of gauge has allowed us to ``regularize'' the
linearized physical metric $\n,\l$, but we expect that non-linear effects will re-introduce some
negative powers of $m^2$ in  $\n,\l$.

\subsubsection{The  interior star solution}

In order to be able to apply our perturbative calculation to the problem of source matching,
we need to formally  consider a star with sufficiently small density, and
sufficiently large radius, so that our assumptions   $\lambda,\mu,\nu \ll 1$
remain satisfied everywhere. We shall give later (after having taken into
account non-linear effects) the precise conditions that the star
characteristics must satisfy. Evidently, such a low density ``star'' is not
physically relevant to the discussion of real stars, but the point of
this subsection is to show in detail how, in principle, one can
match the well-defined exterior  perturbative solution to a unique interior one.

 The interior linearized equations of motion (again for all four
mass  terms) are:
\ba
&&{\lambda' \over R}+{\lambda \over R^2}=-{m^2 \over 2}(\lambda+3\mu+R \mu')+8\pi G 
\rho\\
&&{\nu' \over R}-{\lambda \over R^2}={m^2 \over 2}(\nu+2\mu)+8\pi G P\\
&&\nu'={2\lambda \over R}
\ea
and additionally the continuity equation is:
\be
P'=-{\nu' \over 2}(\rho+P)
\ee
These equations, give a differential equation for $\nu$:
\be
\nu''+\left({2\over R}+{8 \over 3}\pi G(\rho+P)R\right)\nu'-m^2 \nu=8 \pi G\left(2P+{4 
\over 3}\rho\right)
\ee
Let us assume, for the sake of simplicity, that $\rho=const.\equiv \rho_0$ and  $P \ll \rho_0$, {\it i.e.} a constant 
density star of low pressure, so that we can neglect the pressure $P$ in the above
equation. This assumption means that we are considering  a non-compact
start, with $\R \gg  R_S$, {\it i.e.}  $ R_{\odot} \ll (8 \pi G \rho_0)^{-1/2}$.  We can  then safely
neglect the second addendum in the $\nu'$ parentheses. Thus, we have the simple 
differential equation:
\be
\nu''+{2\over R}\nu'-m^2 \nu=8 \pi G{4 \over 3}\rho_0 \label{intnu}
\ee
which differs from the exterior equation \Ref{extnu} above by the source term on the right-hand side (RHS).
Under the same approximations, the other metric functions are
determined in terms of a solution of the interior $\n$ equation by means of:
 \be
\lambda={1 \over 2}R \nu' ~~~ , ~~~ \mu={ \nu'  \over 2 m^2 R}-{\nu \over 2} \label{lamusol}
\ee

The general solution for the $\nu$ function that is regular
 at $R=0$ depends on only one integration constant, say $C_2$ , and reads:
\be
\nu(R)=-{4 \over 3}{8 \pi G\rho_0 \over m^2}+{C_2 \over R}\sinh (mR)
\ee
The continuity of $\nu$ relates the two integration constants, $C_1$ of the exterior 
solution and $C_2$ of the interior one, as:
\be
C_2={R_{\odot} \over\sinh (mR_{\odot}) }\left[{4 \over 3}{8 \pi G\rho_0 \over m^2}-{C_1 
\over R_{\odot}}e^{-m R_{\odot}}\right]
\ee
The continuity of the first derivative of $\nu$, determines  the constant $C_1$ in terms 
of the mass and the radius of the star:
\be
C_1={4R_S \over (mR_{\odot})^3}[mR_{\odot}\cosh (mR_{\odot})-\sinh (mR_{\odot}) ]
\ee
Consistently with our problem of studying the $ m \to 0$ limit, we shall
assume that the radius of the star is such that $ R_{\odot} \ll m^{-1}$. We then get that:
\be
C_1 \approx {4 \over 3}R_S
\ee
[In the opposite limit,   $ m^{-1} \ll R_{\odot}$ we would have obtained that $C_1
\approx{4R_S \over (mR_{\odot})^2}e^{mR_{\odot}}$].
Note that the factor ${4 \over 3}$ here is linked to the well-known fact that,
if one assumes the validity of linearized theory, one must renormalize
the ``bare Newton constant'' $G$ appearing in the massive gravity action by
$  {4 \over 3} G \equiv G_N$ to recover the usual Newtonian constant $G_N$.

Finally, the  full solution is obtained by inserting the matched solution for $\n$
into the expressions \Ref{lamusol} above for the functions $\lambda$  and $\m$.

 Let us finally sketch the structure of the (unique) solution of perturbation theory,
 in the $\l,\m,\n$ gauge, when taking into account the next order in perturbation theory.
 If we introduce $ U \sim e^{-mR}R_S/R \sim R_S/R$ (in the region $\R < R \ll m^{-1}$), we can write:
\ba
 \n \sim \l  \sim  U + \left(1 + { 1 \over mR} + \cdots + { 1 \over (mR)^4} \right) U^2 + \cdots ~,\\
\m  \sim  { U \over (mR)^2}  + \left( { 1 \over (mR)^2} + \cdots + { 1 \over (mR)^6}\right) U^2 + \cdots,
\ea

 This expansion is related to the general perturbation expansions \Ref{expansion}, \Ref{improvedexpansion}  above. In fact, roughly speaking the expansion for the ``physical variables'' $\n,\l$
 corresponds to the gauge-improved expansion   \Ref{improvedexpansion}, while the expansion
 for the gauge variable $\m$ corresponds  to the expansion of the gradient $\de \xi$ of the ``improving''
 gauge transformation  $\xi$ introduced in section 2 above.

 One would need to study more carefully the structure of higher order
 terms in the expansions above to delineate what are the most relevant length scales
 $R_{p,q}$ , see Eq. \Ref{pqscale}, determining the range of validity of
 the  $\l,\m,\n$ perturbation theory, {\it i.e.} for determining the constraints on
 $\R$ and $\r_0$ ensuring that the series above (considered both in the interior
 and the exterior) make sense.  As it is clear that the crucial powers of $mR$ entering
 the successive monomials $ U^n/(mR)^p$ will increase  linearly with the perturbation
 order $n$, there will be a finite limit, when $ n \to \infty$,
 to the sequence of relevant scales  $R_{p,q}$,
 and therefore this sequence will have a finite, global {\it least upper bound},
 or {\it supremum},  $R_{\rm sup}$.  Perturbation theory is then valid (everywhere)
 if $R_{\rm sup} \ll  \R $.

It is evident that since we need (for being able to use perturbation theory)
to constrain ourselves to stars of  extremely small density, the
solution that we have obtained is unrealistic. However, we have gone through this explicit
derivation to show how, in principle,  our various physical requirements (decay at infinity,
matching at the star radius, and regularity at the origin) determine a {\it unique} solution
 in terms of the equation of state ($\rho_0$) and the mass  of the star, and the mass of the graviton.
 Note that this {\it uniqueness} of the solution is a non-trivial consequence of the treatment based
 on perturbation theory. Indeed, perturbation theory allowed not only to provide boundary
 conditions at infinity that selected one solution from the second-order differential
 equation for $\n$,  but it also allowed one to kill the ``sixth degree of freedom'' linked
 (in the $\l,\m,\n$ gauge) to the fact that the equation $f_g =0$ is  a {\it second-order}
 differential equation in $\m$.  The perturbative treatment allowed us to recursively determine
 the higher derivatives  of $\m$ from the algebraic calculation of   $\m$ in terms of
 $\n$ and  $\n'$   (see Eq.    \Ref{lmintermsofn}).

\section{On a claim by Vainshtein}

In the old paper of Vainshtein \cite{Vainshtein:sx}, it was claimed that continuity in the massless
limit could be obtained if one, instead of treating the problem perturbatively, used a different
 ``non-perturbative'' expansion, proceeding in {\it positive} powers of $m^2$,  whose first terms
for the physical variables $\lambda$ and $\nu$ were their usual  GR values, and whose
first term for the gauge variable $\m$ were (in the exterior region) of order:
\be
\label{lmn}
\m \sim \sqrt{-\n} \sim \sqrt{\l} \sim \sqrt{ R_S/R}
\ee
The aim of this section is to critically reexamine this claim, to explain   in detail some points
 which have  remained unclear,  and  to discuss its   serious difficulties.

 A first point to clarify concerns the theoretical framework chosen to ``define'' massive
 gravity. Indeed, the starting point of the original reference \cite{Vainshtein:sx} was not an action, but the
Einstein equation with a postulated energy-momentum tensor for a massive graviton.
However, it is easy to see (by checking the dissymmetry in $(x,y)$ of the  functional
derivative of  the postulated $\sqrt{g(x)} T^{(V)}_ {\m\n}(x)$ with respect to $g^{\a\b}(y)$) that the field
equations postulated in  \cite{Vainshtein:sx} {\it cannot be derived from any action}
(and, in particular, they correspond to none of the four models written
down at the beginning of this paper).
This  makes them quantum mechanically inconsistent.
The energy-momentum tensor postulated in \cite{Vainshtein:sx} reads
(after correcting the sign of the $m^2$ terms as they appear in Eq. (1) there, which actually
corresponded to a tachyonic mass term):
\be
T_{\mu \nu}^{(V)}=-{m^2\over 16\pi G}(h_{\mu \nu}-h \eta_{\mu \nu})\label{VTmn}
\ee

 Note also that, in the same paper, the explicit (t,t) and (R,R) components of the Einstein equations
 appear with the correct sign, consistent with the $T_{\mu \nu}^{(V)}$ given above.
 For this energy momentum tensor the functions $f_t$, $f_R$ and $f_g$ are
given in Appendix A. Note also that in \cite{Vainshtein:sx},  $f_g$ was written down
incorrectly, nevertheless the  final result of \cite{Vainshtein:sx}  starting from Eq. 
(\ref{VTmn}) is mathematically correct.

The basic idea of    \cite{Vainshtein:sx} was the following. The root of the discontinuous
behaviour between  massive gravity and GR is the ``new''  constraint    $f_g$, whose
non-linearity expansion starts with (when keeping terms relevant when Eq. \Ref{lmn} holds):
\be
 \label{mV}
-{1 \over R}f_g =
{\lambda \over R^2}-{\nu' \over 2R}+{4\mu \mu' \over R}+{7\mu'^2 \over
4}+\mu \mu'' + \cdots=0
\ee
The ellipsis in \Ref{mV} contain quadratic terms ${\O}( \m\n + \m\l +  \n^2 + \cdots )$
as well as higher non-linearities.

When this equation is solved in the usual perturbation theory, {\it i.e.}
by solving for the linear terms ${\cal O} (\l)$ and   ${\cal O} (\n')$
and then adding the quadratic terms $  {\cal O} (\m^2 + \m\n + \cdots )$ as corrections,
one gets the ``discontinuous'' result   \Ref{disc} above.
The observation then was that one might cure the discontinuity by assuming that
the     $  {\cal O} (\m^2)$ terms in Eq. \Ref{mV} are comparable to the linear
ones   ${\cal O} (\l)$ and   ${\cal O} (\n')$.  In more detail, one assumes that
the three functions    $\lambda$, $\mu$, $\nu$ admit an expansion in {\it positive}
powers of $m^2$, say:
\be
f(R)=\sum_{n=0}^{\infty} m^{2n}f_n(R)
\ee
where the leading terms,   $\nu_0$, $\lambda_0$, $\mu_0$, are such that
the first two are solutions   of  Eqs. (\ref{laeq}), (\ref{nueq}) with $m^2$    set
to zero, {\it i.e.} (in the exterior):
\be
\label{n0l0}
- \nu_0(R)= \lambda_0(R)= - \ln \left(1-{R_S \over R} \right) =   {R_S \over R} + {1 \over 2}  \left( {R_S \over R}\right)^2
+ \cdots
\ee
while  $\mu_0$    is   a  solution of the constraint equation \Ref{mV} with the above given $\nu_0$, $\lambda_0$
substituted, {\it i.e}.
 \be
 \label{m0V}
{\lambda_0 \over R^2}-{\nu_0' \over 2R}+{4\mu_0\mu_0' \over R}+{7\mu_0'^2 \over
4}+\mu_0\mu_0'' + \cdots =0
\ee
One is looking for a solution $\mu_0$ of Eq. \Ref{m0V} which satisfies (in the exterior region) Eq. \Ref{lmn}
at lowest order. Note that the integration constant associated with $\n_0$ has been set to zero so that
one is consistent  with the ``trivial-vacuum'' (see section \ref{poincarepert}).  Note also that the
quantity $R_S$ entering Eq. \Ref{n0l0} is, at this stage, a GR-like integration constant whose
exact link with the mass of the source is not important.

One then finds that the above requirements lead to a {\it unique} (exterior) solution  for
$\m_0$, which further admits an expansion in powers of $R_S/R$ starting as:
\be
 \mu_{0}=\sqrt{8 \over 13}      \sqrt{R_S \over R}   \left( 1 +  {61 \over 12 \sqrt{26}}  \sqrt{R_S \over R}
 +     {39281 \over 13104} {R_S \over R} + \cdots \right)
 \ee
where the higher order terms are obtained by considering the constraint equation \Ref{m0V} augmented with
higher-order terms in  $\l_0$, $\n_0$ and $\mu_0$.

 Starting from this result for      $\mu_{0}$ (where all higher-order
 coefficients are, in principle, uniquely determined), we then go back to Eq.  (\ref{laeq}) (in the exterior region),
 which yields a linear first-order differential equation for  the ${\cal O}(m^2)$ term $\lambda_1$
in $\l$:
\be
\left({1 \over R^2} +\cdots\right)\l_1+\left({1 \over R}+\cdots\right)\l'_1 =-{5 \over \sqrt{26}} \sqrt{R_S \over R} -   {257 \over 156} {R_S \over R}+\cdots
\ee
The general solution of this inhomogeneous linear ordinary differential equation (ODE) can be written as the sum of a
particular inhomogeneous solution:
\be
m^2 \l_1 = (m R)^2 \left(-\sqrt{2 \over 13}   \sqrt{R_S \over R}    - {257 \over 312} {R_S \over R}  + \cdots \right)
\ee
plus the general solution of the homogeneous equation, which  is of the form
$c_{\l1}/R$.  The latter general homogeneous solution can be absorbed in a
 ${\cal O}(m^2)$ change of the integration constant $R_S$, and can therefore be physically ignored.

We can then insert the known values of    $\nu_0$, $\lambda_0$,   $\mu_0$  and    $\lambda_1$
in Eq. \Ref{nueq} to get a  linear first-order differential equation for  the ${\cal O}(m^2)$ term $\n_1$
in $\n$:
\be
\left({1 \over R}+\cdots\right)\n'_1 = \sqrt{2 \over 13} \sqrt{R_S \over R} - {13 \over 24}   {R_S \over R}+\cdots
\ee
Again,  the  general solution of this inhomogeneous linear ODE can be written as the sum of a
particular inhomogeneous solution:
 \be
m^2 \n_1 = (m R)^2 \left(  {2 \over 3}\sqrt{2 \over 13} \sqrt{R_S \over R} -{13 \over 24}    {R_S \over R}  + \cdots \right)
\ee
plus the general solution of the homogeneous equation, which  is simply a constant $c_{\n1}$.
The latter constant can be absorbed in a rescaling of the time variable, and can
therefore be physically ignored.

The next step is more tricky.  Indeed, we get a second-order linear ODE  for
the  ${\cal O}(m^2)$ term $\m_1 $ in $m$  of the type:
\be
\label{m1V}
(\m_0 + \cdots)\m_1'' + \left( 2 a \m_0' + {b \over R}\m_0 + \cdots\right) \m_1'  + \left(\m_0'' + {b \over R} \m_0'+ \cdots\right) \m_1 =
   {3 \over \sqrt{26}}\sqrt{R_S \over R} + {813 \over 624}   {R_S \over R}  + \cdots
\ee
Here $ a = 7/4 $ and $b=4$ are the coefficients appearing in the $\m$-quadratic terms
in the constraint   \Ref{mV}. The problem now is that the general homogeneous solution
of  Eq.  \Ref{m1V}  a priori introduces two new integration constants $c_{m1+}$, $c_{m1-}$
in a solution of the type $ \m_1^{\rm hom} =  c_{\m1+} R^{s_+} + c_{\m1-}R^{s_-}$ where the exponents
$ s_+$,  $ s_-$  are the two roots of the quadratic indicial equation
$ s^2 + (b-a-1) s + 3/4 - b/2 =0~ \Rightarrow~s_{\pm}={\sqrt{5} \over 8}(-\sqrt{5}\pm\sqrt{21})$.    Contrary to what happened above for the
integration constants entering  the homogeneous solutions  in $\l_1$ and $\n_1$,
the $\m_1$-integration constants cannot a priori  be physically ignored.  However, if
we follow the spirit of  \cite{Vainshtein:sx} and of the leading requirement   \Ref{lmn}, {\it i.e.} if we try to construct the simplest vacuum solution which is entirely determined
by the GR-like integration constant $R_S$,  we can discard the above homogeneous
solution and continue the iteration by selecting for $\m_1$ the particular
 solution    determined by the inhomogeneous ``source terms'' in the   $\m_1$-equation.
 This solution has the following solution:
 \be
 m^2 \m_1 = (m R)^2 \left( {1 \over 7} +  {1305 \over 644 \sqrt{26}} \sqrt{R_S \over R}  + \cdots\right)
 \ee
 Note that the exponents  $ s_+$,  $ s_-$    entering the $\m_1$-homogeneous solution
 are irrational.  Taking into account the possibility of adding
 this homogeneous solution would introduce a new sequence of characteristic
 length scales \Ref{pqscale} which would mix in a complicated way with the other
 (rational) scales.

 Continuing in the same way  (always  discarding the homogeneous solutions
 entering the $\m$ equations)  we end up with a double
series  for the three functions $ f= \n,\l,\m$ of the form:
\be
f= \left( {R_S \over R} \right)^{a} \sum_{n=0}^{\infty} \sum_{k=0}^{\infty}  f_{nk}\left( {m^2
R^{5/2} \over {R_S}^{1/2}} \right)^{n}\left( {R_S}^{1/2} \over R^{1/2} \right)^k
\label{Vexp}
\ee
where $a=1$ for $\lambda$, $\nu$ and $a=1/2$ for $\mu$.
For notational simplicity, we have not indicated in Eq. \Ref{Vexp} the presence
of the {\it logarithmic} terms $\log (R/S)$  which appear each time the ``source terms''
for the perturbations in $\l$ and $\n$
contain a power of $R$ which match (modulo a factor $R^2$) the power of
$R$ of a homogeneous solution of the left-hand-side. The arbitrary scale  $S$  that
one can introduce in these logs is not physically important because a rescaling
of $S$ corresponds to the addition of an (ignorable) homogeneous solution  in    $\l$ and $\n$.
[Because of the irrational character of   $ s_+$,  $ s_-$, no logs primitively
enter from the perturbation equations for the $\m$ variable.]
Without loss of generality, we can then choose $S=R_S$ as scale in all the logs.

The final result \Ref{Vexp} thereby involves only two length scales  $R_S$ and the
length scale entering the monomial    ${m^2 R^{5/2} \over {R_S}^{1/2}} $.
For this expansion to make sense,
we should make sure that both ${R_S \over R}$ and  ${m^2 R^{5/2} \over {R_S}^{1/2}} $
be (much)  smaller than  1. This requirement is 
satisfied if we look at distances $R_S \ll R \ll R_V$, where $R_V$ is the length
scale introduced in Eq. \Ref{Vscale}, which, using the general definition \Ref{pqscale}
can be written as:
\be
\label{Vscale'}
R_V= R_{4 , 1}
\ee

To summarize so far: we have verified part of the claim of   \cite{Vainshtein:sx}, namely
the existence of a (particular)  double expansion, a priori valid in the intermediate
range  $R_S \ll R \ll R_V$, that (i) represents {\it a}   formal  (vacuum) {\it local} solution  of the Einstein
equations modified by the mass  term  \Ref{VTmn}, (ii) is continuous as $m^2 \to 0$,
and (iii) satisfies Eq. \Ref{lmn}.

The weak points of this argument are several. First, the specific local solution chosen in Eq.  \Ref{Vexp}
corresponds to a postulated
energy momentum tensor which does not stem from an action, and we
are going to see that a similar solution {\it does not exist} for the action used as a 
starting point in \cite{Deffayet:2001uk}.
But the weakest point concerning the  expansion   \Ref{Vexp}
is the lack of  rationale indicating that this local solution does approximate the
unique global one satisfying  the standard boundary conditions at infinity \Ref{bc}
and being regular at the origin.  The detailed argument above, showing that two
arbitrary integration constants have been set to zero to obtain Eq. \Ref{Vexp},
does not make it a priori probable that Eq. \Ref{Vexp} happens to satisfy all
the required boundary conditions.

Before tackling the more complicated ``global'' issues, let us first study the issue
of the existence of formal Vainshtein-type expansions for {\it action-based}
field equations.   We start by considering the potential
${\cal  V}^{(1)}$  that appeared in the recent paper of
\cite{Deffayet:2001uk}, which re-discussed the result of \cite{Vainshtein:sx}.
Trying to repeat the same procedure
as above we face the following problem. The equation for $\mu_0$ now reads:
\be
{\lambda_0 \over R^2}-{\nu_0' \over 2R}-{6\mu_0\mu_0' \over R}-{3\mu_0'^2 \over
4}-{3\mu_0\mu_0'' \over 2} + \cdots =0
\ee
The $\m$-quadratic terms in this equation are different from  those in Eq. (\ref{m0V}).
This difference is crucial because if we look again for a solution satisfying   Eq.  \Ref{lmn}, {\it i.e.} admitting an expansion of the form
$\m_0 = \m_{00} \sqrt{R_S/R} + \m_{01} (R_S/R) + \cdots$ we find that there
exists no such solution\footnote{This contradicts the statements of \cite{Deffayet:2001uk},
which were supposed to concern the mass term ${\cal  V}^{(1)}$. They, instead,
seem to correspond to the  postulated
$T^{(V)}_{\mu\nu}$ of \cite{Vainshtein:sx}.}.
[Formally, the only such solution starts with
an imaginary coefficient:     $\mu_{00}=\pm i \sqrt{{8 \over 27}}$.]

The same
problem persists if we consider the potential ${\cal  V}^{(2)}$. Then the same equation 
reads:
\be
{\lambda_0 \over R^2}-{\nu_0' \over 2R}-{2\mu_0\mu_0' \over R}-{\mu_0'^2 \over 
4}-{\mu_0\mu_0'' \over 2} + \cdots=0
\ee
and would again formally  give an imaginary solution: $\mu_{00}=\pm i \sqrt{{8 \over 9}}$.

The situation is better, however, for the potentials ${\cal  V}^{(3)}$ and ${\cal
V}^{(4)}$.   The approximated
equation for  ${\cal  V}^{(3)}$ is then:
\be
{\lambda_0 \over R^2}-{\nu_0' \over 2R}+{2\mu_0\mu_0' \over R}+{\mu_0'^2 \over 
4}+{\mu_0\mu_0'' \over 2} + \cdots=0
\ee
which gives $\mu_{00}=\sqrt{{8 \over 9}}$ and for ${\cal  V}^{(4)}$ we have:
\be
{\lambda_0 \over R^2}-{\nu_0' \over 2R}+{6\mu_0\mu_0' \over R}+{3\mu_0'^2 \over 
4}+{3\mu_0\mu_0'' \over 2} + \cdots=0
\ee
which gives $\mu_{00}=\sqrt{{8 \over 27}}$. Thus, at least in these cases,  a Vainshtein-type
expansion similar to \Ref{Vexp}  exists. Again, for these expansions to make
sense one should be at the intermediate range of distances $R_S \ll R \ll R_V$,
with $R_V$ given by Eq. (\ref{Vscale}).

Let us note in passing that the  new  scale selected by a Vainshtein-type construction
 is not at all universal for different mass
terms. For example, let us consider the potential \cite{Damour:2002ws}:
\be
{\cal V}^{(\sigma)}=(f g)^{1/4}(\sigma_2-\sigma_1^2) \label{Vs}
\ee
where  $\s_n \equiv  \S_a (\ln \l_a)^n $, with $\l_a$ denoting the eigenvalues of ${\bf f}^{-1} {\bf g}$
(see below). Then,    the non-linearity expansion
of the constraint \Ref{coneq} gives:
\be
-{1 \over R} f_g =  {\l \over R^2}-{\n' \over 2R}+{3 \m^2 \m' \over 8R }+{\m \m'^2 \over 4}+{R \m'^3 \over 48}+{3\m^2 \m'' \over 32}+{R\m \m' \m'' \over 16} +\cdots =0
\ee

If we introduce the short-hand notations $\ep \equiv  m R_S$ and $x\equiv R/R_S$, 
we end up with instead of Eq. (\ref{Vexp}), expansions of the type:

\be
f=x^{-a}\sum_{n=0}^{\infty} \sum_{k=0}^{\infty}  f_{nk}\left(\epsilon^2 
x^{8/3}\right)^{n}\left(x^{-1/3}\right)^k
\ee
with  $a=1$ for $\lambda$, $\nu$ and $a=1/3$ for $\mu$. This expansion has the same form 
even if we add the stabilizing $\lambda \sigma_2^2$ term (see \cite{Damour:2002wu}). By 
the same argument as before, this gives the following range of distances for which the  above
expansion makes sense:
\be
R_S \ll R \ll {R_S \over (mR_S)^{3/4}}  \equiv R_{3,1}
\ee

A final example would be to consider the following potential:
\be
{\cal V}^{(0)}=(f g)^{1/4}\sigma_2^2 \label{V0}
\ee
for which  the mass of the graviton is zero, but still we have a non-trivial potential 
which breaks general covariance. Then the expansion at lowest order looks like:
\be
f=x^{-1}\left[f_{00}+f_{10}\epsilon^2 \log x+\cdots \right]
\ee
for all three functions $\mu$, $\nu$ and $\lambda$. The formal range of validity of this
expansion is now:
\be
R_S \ll R \ll R_S  e^{1/ (mR_S)^2}
\ee

At this point, it is important to remark that, contrary to what one might think (in view of the
controversy over the ``discontinuity'' issue) {\it there exist many  local solutions }
of the massive-gravity field equations which are continuous as $m^2 \to 0$ and which,
therefore, might suggest that massive gravity is compatible with local tests of GR.
The specific Vainshtein construction is   (when it works) one way of exhibiting such
solutions (constrained to the obtention of strictly stationary solutions).
However, we have seen above that this specific way fails in important cases,
such as the simplest mass term         $ {\cal V}^{(1)}$.  However, even when there does
not exist a $\m_0$ satisfying the simple requirement \Ref{lmn}, there might still exist other
solutions. Indeed, the  equation that  $\m_0$   has to satisfy  is a second-order
 quasi-linear ODE  and one can always construct local solutions (depending on
 two arbitrary parameters) for such an equation. [The problem then is to show that
 the domain of existence of the solution is large enough to cover the phenomenological tests.]
 In fact, if we forget for a moment about the restriction to  strictly stationary solutions,
  Ref. \cite{Damour:2002ws}  has indicated how to construct general classes of
  solutions, evolving on the time scale     $m^{-1}$ (thought of as the Hubble time scale),
  which are ${\cal O}(m^2)$ close to any desired GR solution   everywhere in space.
  All these constructions leave open, however, two basic issues:
  (i) can these local solutions be extended to globally asymptotically flat  solutions,
  or, if they cannot, (ii) can they be extended to a cosmological solution which is a natural
  ``attractor'' of the cosmological dynamics, so that it is indeed natural to use
  such a solution to describe the universe around us?

Before tackling the issue of whether the formal, particular  Vainshtein-type  expansions
are part of a globally regular solution (both at the origin and at infinity), we study in the
next section the structure of the field equations in a different gauge.

\section{The $c$, $\bar{c}$, $\bar{b}$ variables}

Let us consider the  $a,b,c$ gauge \Ref{abc}, and replace the three basic variables $a,b,c$
by the equivalent combinations  $c$, $\bar{c}$, $\bar{b}$ where we define:
\be
 \bar{b}=ab \, \,  \, , \,  \,  \,   \, \bar{c}=ca^2
 \ee

Let us indicate the link between the $\lambda$, $\mu$, $\nu$ variables and the
 $a,b,c$ ones (or  the equivalent $c$, $\bar{c}$, $\bar{b}$ ones).
 We first  compute from the three functions $a(r)$, $b(r)$, $c(r)$ the functions
  $\lambda(r)$, $\mu(r)$, $\nu(r)$ as follows:
\be
a^2(r)=e^{\nu(r)}~~~,~~~b^2(r)=e^{\lambda(r)+\mu(r)}\left(1+{r \over 2}{d\mu \over
dr}\right)^2~~~,~~~c^2(r)=r^2 e^{\mu(r)} \label{conn}
\ee
Then  we make the coordinate transformation $ r \to R$ by inverting  $R= c(r) = r e^{\mu(r)/2}$.

 The advantage of introducing these variables is that they simplify the Lagrangian
 formulation of the dynamics.
 Indeed, the action can easily be checked to be:
\be
{\cal S}={1 \over 2 G}\int dt dr  \left({\dot{c}~\dot{\bar{c}} \over
\bar{b}}+\bar{b}\right)- {m^2 \over 16 G} \int dt dr \bar{{\cal V}}(c,\bar{c}, \bar{b},r)+{\cal S}_{{\rm matt}}
\ee
where we have introduced  $\bar{{\cal V}} \equiv { \cal V} / \sin \theta$.
Note that  ${\cal V}$ has the scaling behaviour
${\cal V} = r^2 f(c/r,\bar{c}/r,\bar{b})$.
 We have denoted $\dot{}\equiv d /dr $.  The pair $(c,\bar{c})$ appears here  as ``light-cone'' coordinates for a
``relativistic particle''.   We refer to the cosmological studies
of massive gravity\cite{Damour:2002ws,Damour:2002wu}
for a discussion of similar ``relativistic particle Lagrangians''.  Note that the variable $\bar{b}$
is a ``gauge-like'' variable ({\it radial lapse}) which does not have   a kinetic term, just as the
(relative) {\it time lapse} variable  $e^{\gamma}$ did  in cosmological studies. Therefore the
equation of motion for  $\bar{b}$ is an algebraic equation, instead of the second-order ODE's
one gets for the dynamical variables   $(c,\bar{c})$.

The equations of motion for this gauge choice are:
 \ba
&&{1 \over  \bar{b}} { d \over dr} \left({1 \over  \bar{b}}  { d c \over dr} \right)
=-{m^2 \over 8}{1 \over \bar{b}}{\de \bar{{\cal V}} \over  \de
\bar{c}}-4 \pi G {c^2 \over \bar{c}}(\rho+P_r)\label{rceq}\\
&&{1 \over  \bar{b}} { d \over dr} \left({1 \over  \bar{b}}  { d\bar{c} \over dr} \right)
=-{m^2 \over 8}{1 \over \bar{b}}{\de \bar{{\cal V}}
\over  \de c}+4 \pi G c (\rho+P_r+4P_t)\label{rcbeq}\\
&&{1 \over  \bar{b}^2} { d c \over dr}{d \bar{c} \over dr}-1=-{m^2 \over 8}{\de \bar{{\cal V}}
\over  \de \bar{b}}+8 \pi G c^2 P_r\label{rbeq}
\ea
A consequence of these equations are  the conservation equation of the matter
energy-momentum tensor:
\be
\label{Peq}
2 {dP_r \over dr}+{d \log c \over dr}(3P_r-\rho-4P_t)+{d \log \bar{c}
\over dr}(\rho+P_r)=0
\ee

Note that the structure of the equations of motion in vacuum ($\rho = P_r= P_t =0$)
is relatively simple: Eq. \Ref{rbeq} is an algebraic equation for  $ \bar{b}$ whose
solution yields $    \bar{b} = B(dc/dr, d  \bar{c}/dr, c,  \bar{c} , r)$.  Inserting this expression
in the first two equations then yields two second-order, non-linear ODE's for the radial
evolution  of $c, \bar{c}$.

We note that it is sometimes convenient to reformulate the field equations by changing
the radial variable from $r$ to the radial analog of the ``proper time'', namely
$  \tilde{r}$ defined by $ d  \tilde{r} \equiv  \bar{b} dr$. Then the set of field
equations read:
\ba
&&{d^2 c \over d\tilde{r}^2}=-{m^2 \over 8}{1 \over \bar{b}}{\de \bar{{\cal V}} \over  \de
\bar{c}}-4 \pi G {c^2 \over \bar{c}}(\rho+P_r)\label{ceq}\\
&&{d^2 \bar{c} \over d\tilde{r}^2}=-{m^2 \over 8}{1 \over \bar{b}}{\de \bar{{\cal V}}
\over  \de c}+4 \pi G c (\rho+P_r+4P_t)\label{cbeq}\\
&&{d c \over d\tilde{r}}{d \bar{c} \over d\tilde{r}}-1=-{m^2 \over 8}{\de \bar{{\cal V}}
\over  \de \bar{b}}+8 \pi G c^2 P_r\label{beq}\\
&&{dr \over d\tilde{r}}={1 \over \bar{b}}\label{rtdef}
\ea
 Note  the
appearance of a fourth equation in the system linked to the definition of $  \tilde{r}$ .
 The pressure-balance equation   takes the same form as Eq. \Ref{Peq} with the
 replacement $dr \to d \tilde{r}$.

The use of  the variable $\tilde{r}$ corresponds to writing the two metrics as:
\ba
&&ds^2=-a^2(\tilde{r})dt^2+{d\tilde{r}^2 \over a^2(\tilde{r})}+c^2(\tilde{r}) d\Omega^2\\
&&ds_{\rm fl}^2=-dt^2 +{d\tilde{r}^2 \over \bar{b}^2(\tilde{r})}+r^2(\tilde{r}) d\Omega^2
\ea

\subsection{The General Relativity $m=0$ limit}

To get some familiarity with the     $c$, $\bar{c}$, $\bar{b}$ variables let us see how one
can derive the Schwarzschild solution in this  new language.  In GR   $\bar{b}$ is a gauge
variable, and we can set    \mbox{$\bar{b}=1$} for the exterior solution which implies that,
after fixing an integration constant, we can write $\tilde{r}=r$. Alternatively, we can start from the $\tilde{r}$-formulation
and set to zero the terms ${\cal O}(m^2)$. This yields the very simple evolution
equations   $d^2 c / d\tilde{r}^2 = 0 =  d^2  \bar{c } / d\tilde{r}^2$, submitted to the
constraint      $( d c / d\tilde{r}) ( d  \bar{c } / d\tilde{r}) = 1$. The general solution is
$ c = \a   \tilde{r} + \b,    \bar{c } = \g  \tilde{r} + \d$, with the constraint $\a \g =1$.
The solution depends on three arbitrary constants. However, we can use the remaining rigid gauge symmetries
of the problem ($ \tilde{r}' = a \tilde{r} + b,  t' = c t$) to set $\a$ to 1 and $\b$ to zero.
This leaves only one physical integration constant,    $ \d  \equiv -R_S$, in terms of which
the solution finally reads:
\be
\label{schw}
c=\tilde{r}= r~~~~,~~~~\bar{c}=\tilde{r}-R_S = r -R_S
\ee

For the interior solution, and for a constant density star, we choose the gauge variable to have the form:
\be
\bar{b}={3 \over 2}\sqrt{1-{R_S \over R_{\odot}} \over 1-{R_S \over R_{\odot}^3}r^2}-{1 
\over 2}
\ee
so that we continue having  $c=r$. Then the $\bar{c}$ function reads: 
\be
\bar{c}=r\left({3 \over 2}\sqrt{1-{R_S \over R_{\odot}}}-{1 \over 2}\sqrt{1-{R_S \over 
R_{\odot}^3}r^2}\right)^2
\ee
Note that the interior solutions can be obtained from the ones of the  $\lambda$, $\mu$, $\nu$ gauge [Eqs. \Ref{GRintla}, \Ref{GRintnu}], using the connecting relations (\ref{conn}) with $\m=0$.

\subsection{The perturbative limit for $m \neq 0$}
\label{pertsolc}

We can look again how the results change in the  linearized approximation for the massive
case  in terms of the $c$, $\bar{c}$, $\bar{b}$. In this case we linearize as:
\be
c=r+\delta c ~~~,~~~ \bar{c}=r+\delta\bar{c} ~~~,~~~\bar{b}=1+\delta\bar{b}
\ee
We will only present here the exterior star solution to verify the correspondence with the 
$\lambda$, $\mu$, $\nu$ gauge. The vacuum equations (\ref{ceq}), (\ref{cbeq}), 
(\ref{beq}), (\ref{rtdef}), using the $r$ variable become:
\ba
&&\delta c'' -\delta \bar{b}'={m^2 \over 2}(r\delta \bar{b}+\delta c-\delta \bar{c} )\label{dceq}\\
&&\delta \bar{c}'' -\delta \bar{b}'={m^2 \over 2}(3r\delta \bar{b}+3\delta c+\delta 
\bar{c} )\label{dcbeq}\\
&&\delta c'+\delta \bar{c}'-2 \delta \bar{b}={m^2 \over 2}r(3\delta c+\delta \bar{c})\label{dbeq}
\ea
for all four initial potentials (\ref{V1}), (\ref{V2}), (\ref{V3}), (\ref{V4}). From Eq. \Ref{dbeq} we can solve for $\delta \bar{b}$ and substitute back to Eqs. \Ref{dceq}, \Ref{dcbeq}. Adding the resulting equations we obtain:
\be
r(\delta c'-\delta \bar{c}')+(1+m^2r^2)\delta \bar{c}-(1-3m^2 r^2)\delta c=0\label{new1}
\ee
Differentiating the above equation and substituting the difference $\delta c''-\delta \bar{c}''$ from Eq. \Ref{dceq} (with $\delta \bar{b}$ substituted), we obtain a second independent first order equation:
\be
r(2\delta c'+\delta \bar{c}')+{1 \over 4}(2-m^2 r^2)\delta \bar{c}+{1 \over 4}(22-3m^2r^2)\delta c=0\label{new2}
\ee
From Eqs. \Ref{new1}, \Ref{new2} we can solve for $\delta\bar{c}$ and substitute to any of them. In this way we get a second order differential equation for $\delta c$:
\be
\delta c''+{4 \over r(2+m^2r^2)}\delta c'-{2+m^2r^2 \over r^2}\delta c=0 \label{finaldc}
\ee
while $\delta\bar{c}$ and $\delta\bar{b}$ are then given as:
\be
\delta\bar{c}=-3\delta c-{4r \over 2+m^2r^2}\delta c'~~~, ~~~
\delta \bar{b}={1 \over 2}\delta c'+\delta \bar{c}'-{m^2 \over 4}r(3\delta c+\delta \bar{c})
\ee

Thus, although the system \Ref{dceq}, \Ref{dcbeq}, \Ref{dbeq} at first sight seems to involve four integration constants, in fact only two of them are independent. Finally, we can integrate Eq. \Ref{finaldc} and write the general solution decaying at infinity  as:
\ba
&&\delta c={C_1 \over 4}\left(1+{1 \over mr}+{1 \over (mr)^2}\right)e^{-mr}\label{solinftyc}\\
&&\delta \bar{c}={C_1 \over 4}\left(-3+{1 \over mr}+{1 \over (mr)^2}\right)e^{-mr}\label{solinftycb}\\
&&\delta \bar{b}=-{C_1 \over 2r}\left(1+{1 \over mr}+{1 \over (mr)^2}\right)e^{-mr}\label{solinftybb}
\ea
where we have set the second integration constant corresponding to the exponentially growing solution to zero.
It is easy to see, from the connecting relations (\ref{conn}) and after defining $R=r
e^{\mu/2}$, that these solutions are identical with the ones obtained in the $\lambda$, 
$\mu$, $\nu$ gauge [Eqs. \Ref{nupert}, \Ref{lapert}, \Ref{mupert}], with the same integration constant   $ C_1$. The latter, as we saw in section \ref{pertsol}, for $ R_{\odot} \ll m^{-1}$ is $C_1 \approx {4 \over 3}R_S$.

\section{Singularity in the general asymptotically flat solution}

Let us now discuss the central question of whether non-linear effects  do cure or not
the ``discontinuity'' (as $m^2 \to 0$) with respect to GR exhibited by the
linearized approximation.  We recall that the specific conjecture made by Vainshtein
in this respect was that the  non-linear ``dressing'' of the unique perturbative solution (defined
by resumming Poincar\'e-covariant diagrams, as sketched in \cite{Deffayet:2001uk})
leads to a solution   of the type constructed in  section 5.
 There are several ways one can take to investigate this conjecture. Either one can start from the
center of the star with the requirement that it is locally flat and integrate outwards, or
start from ``infinity'' (practically from some distance $\gg m^{-1}$) with the requirement
of asymptotic flatness and integrate inwards. In addition, we can work in several different gauges.

Our work had to be necessarily numerical because it is a difficult dynamical question
to control the global structure of  a non-linear theory such as massive gravity.
Though we have done simulations both ways (inwards and outwards), and with several
gauges   ($\lambda$, $\mu$, $\nu$, and $a,b,c$), we shall mainly report on the
results obtained by {\it integrating inwards, in the   $a,b,c$ gauge}.
This way of proceeding has indeed two advantages: (i) the solution is essentially unique
 (once the equation of state of the matter is fixed)
 in the sense that its ``initial state'' at infinity  is physically determined by giving oneself only {\it one}
dimensionless parameter, $mR_S$, and (ii) the (inwards) radial evolution of the convenient
variables $c$, $\bar{c}$, $\bar{b}$ is similar to the cosmological evolution studied in
\cite{Damour:2002wu}, which allows us to draw some intuition about its qualitative
behaviour. Note that, by  contrast, starting the evolution from the center obliges one
to consider a {\it two parameter} family of solutions. One then needs to ``shoot'' from
the center with a one-parameter family of initial data until one eventually matches
the unique exponentially decaying  solution at infinity.

Let us start by discussing the exterior region. [As we shall see this suffices to
conclude negatively.] The dynamics of the   $c$, $\bar{c}$, $\bar{b}$ variables
is given by the Lagrangian:
 \be
 \label{Lag}
{\cal L}={1 \over 2 G} \left({\dot{c}~ \dot{\bar{c}} \over \bar{b}}+\bar{b}\right)- {m^2
\over 16 G} \bar{{\cal V}}\left(\bar{b},c ,\bar{c} , r \right)
\ee

We start at infinity with the solution   \Ref{solinftyc}, \Ref{solinftycb}, \Ref{solinftybb}. For this solution, the ``kinetic   term''
 $ \dot{c}~ \dot{\bar{c}} $ is {\it positive} (and near + 1), which means that the
 analogous ``relativistic particle'' initially moves  on a {\it spacelike} worldline.
 Let us recall that, as $\bar{b}$  has no kinetic term, the value of $\bar{b}$ at any
 ``moment'' ({\it i.e.} at any radius $r$) is determined by extremizing the Lagrangian
\Ref{Lag} considered as a function of  $\bar{b}$.
 As long as the sign  $ \dot{c}~ \dot{\bar{c}} > 0$ does not change, the first two
 terms in the Lagrangian \Ref{Lag} would suffice to ``confine''  $\bar{b}$ around a unique
 extremum $\sim \sqrt{\dot{c}~ \dot{\bar{c}}}$. If ever the dynamics
tends to decrease the value of $ \dot{c}~ \dot{\bar{c}}$, this will tend to drive
 $\bar{b}$ towards smaller values.  In fact the dynamics is due to the interplay between
 the ``kinetic terms'' in \Ref{Lag} and the potential terms and we must also take into account the
 $\bar{b}-$dependence of $\bar{{\cal V}}\left(\bar{b},c ,\bar{c} , r \right)$.
 This dependence is somewhat complicated, and has  a form which depends on the
 specific mass term considered.  For our four fiducial mass terms ${\cal V}^{(i)}$, $i=1,2,3,4$
 we have:
\ba
&&\bar{{\cal V}}^{(1)}=\bar{b}^3F^{(1)}+\bar{b}G^{(1)}\label{Vb1}\\
&&\bar{{\cal V}}^{(2)}=\bar{b}^2F^{(2)}+G^{(2)}\label{Vb2}\\
&&\bar{{\cal V}}^{(3)}={1 \over \bar{b}}F^{(3)}+\bar{b}G^{(3)}\label{Vb3}\\
&&\bar{{\cal V}}^{(4)}={1 \over \bar{b}^2}F^{(4)}+G^{(4)}\label{Vb4}
\ea
where, $F^{(i)}=r^2 F^{(i)}\left({c \over r},{\bar{c} \over r}\right)$,
$G^{(i)}=r^2 G^{(i)}\left({c \over r},{\bar{c} \over r}\right)$.  However, this structure is not
enough for concluding about the ``confining properties''  of the potential terms,
because the coefficients of the various powers of $\bar{b}$ above do not have
 a definite sign. Indeed, the explicit form of these functions is:
\ba
&&F^{(1)}={c^2 \over r^2}F^{(2)}=c^2\left( -12 + 12{c^2 \over r^2}-2{c^4 \over r^4} +6 {\bar{c} \over c} -4{c \bar{c} \over r^2}   \right)\\
&&G^{(1)}={c^2 \over r^2}G^{(2)}=c^2\left(  -2+6{c \over \bar{c}}-4{c^3 \over r^2 \bar{c}}  \right)\\
&&F^{(3)}={c^2 \over r^2}F^{(4)}=c^2\left(-12+ 12{r^2 \over c^2}-2{r^4 \over c^4} +6 {c \over \bar{c}} -4{r^2 \over c \bar{c}} \right)\\
&&G^{(3)}={c^2 \over r^2}G^{(4)}=c^2\left(  -2+6{\bar{c} \over c}-4{ r^2 \bar{c} \over c^3}  \right)
\ea
We can, however,  have an  analytical idea of the
 natural tendency of the dynamics by looking at the linearized solution \Ref{solinftyc}, \Ref{solinftycb}, \Ref{solinftybb}
 which results from the combined effect of kinetic and potential terms.
 If we look at Eq.  \Ref{solinftybb} (remembering that $C_1 >0$)
 we see that all the terms in the variation of $\bar{b}$ are {\it negative}, {\it i.e.} that
  $\bar{b}$ tends to decrease from its initial value  $\bar{b} = 1$.   In addition  one
  can deduce from  Eqs. \Ref{solinftyc}, \Ref{solinftycb}  that the crucial quantity    $ \dot{c}~ \dot{\bar{c}}$,
  after increasing, starts decreasing when $ r  \ltsim m^{-1}$.
  As we said above, such a decrease of    $ \dot{c}~ \dot{\bar{c}}$  tends
to further drive      $\bar{b}$ down.

If we first consider the mass terms  $i=1,2$, we see that they contain only
{\it positive} powers of   $\bar{b}$.  Whatever be the sign of the coefficients
of these terms, such functions are quite inefficient in preventing   $\bar{b}$
from decreasing all the way towards zero, if the tendency of the ``confining''
term    $\dot{c}~\dot{\bar{c}}/ \bar{b}$   in \Ref{Lag} is to drive
 $\bar{b}$ there because    $\dot{c}~\dot{\bar{c}}$ happens to decrease below
 zero.  Therefore, in the cases    $i=1,2$ it  might a priori happen that the
 dynamics drives   $\bar{b}$   toward zero at some finite radius, similarly
 to what we found in the cosmological study      \cite{Damour:2002wu}
 where mass terms with weakly confining features tended to drive
 the gauge function $e^\g$ to zero in a finite time.
 The numerical simulations we performed for the potentials   $i=1,2$,
 with various values of $mR_S$ (in the range $10^{-5} \leq mR_S \leq 10^{-1}$)
 have shown that indeed such a behaviour is generic: we systematically
 find that the inward radial evolution ends up in a singularity\footnote{
 Within the context of massive gravity the value $\bar{b} =0$ corresponds
 to an invariant singularity of the bi-gravity configuration $({\bf g}   ,  { \bf f})$
 because one of the eigenvalues of $ { \bf f}^{-1}  {\bf g} $ is  $ 1/\bar{b}$.}
 where the variable    $\bar{b}$   tends to zero, while at the same time $\dot{c}~\dot{\bar{c}} \to 0$.

 We have also run similar simulations  with the other mass terms $i=3,4$, and we found
 the same singular behaviour:     $\bar{b} \to 0$  at a finite radius, while, at the same time,
  $\dot{c}~\dot{\bar{c}} \to 0$.  In the latter cases, the apparent ``potential barriers''
$ F^{(3)}/  \bar{b}$ and $F^{(4)}/ \bar{b}^2 $ that might have helped to prevent
  $\bar{b}$ to tend to zero turn out to be ineffective, because the (non-positive definite)
  coefficients are found to be driven by the dynamics towards zero at the same time
  that  $\dot{c}~\dot{\bar{c}} \to 0$.

  Our final conclusion (based on our numerical simulations) is therefore that for
  all values $10^{-5} \leq mR_S \leq 10^{-1} $ (and probably also for all
  smaller values $m R_S < 10^{-5}$) the unique, asymptotically flat SSS solution
  (which is well defined near infinity) develops a singularity at some finite
  radius. The latter radius depends on the mass potential and is roughly between $R_p$ and $ m^{-1}$.

 To complement our numerical study, we also performed simulations (still in the $a,b,c$ gauge)
 that start at the center of a star.  This case is more delicate because generic (regular)
 initial data at $r=0$ depend now on {\it two} arbitrary parameters, and we expect that
 only a ``line'' of data can evolve into an asymptotically flat solution.
 If we consider a weakly self-gravitating star, it is relatively easy to choose
 initial data at the center (with  $\bar{b},  \dot{c}$ and $\dot{\bar{c}} $  all near one)
 that satisfy the constraint \Ref{rbeq}.  Then, when considering the
 evolution system with respect to the ``proper radius''  $  \tilde{r}$,  general theorems on the continuity
 (with respect to small parameters appearing  in the coefficients) of solutions of ODE systems
 guarantee that, if $m^2 \R^2 $ is small enough, any regular
interior solution in the GR limit    will be smoothly deformed into some
massive gravity interior one.   This provides a way to exhibit  classes of {\it local} solutions
of massive gravity which are continuously  connected to GR solutions.
However, a hard dynamical question (which is not covered by the local continuity
theorems for ODE's) is to know on which length scale these solutions stay
close to GR solutions, and what are their asymptotic behaviour at infinity.
We have addressed this dynamical question by numerical simulations
(in the $a,b,c$ gauge).  We found that    all our
 simulations evolved a singularity at a finite radius (where   $\bar{b}$ run away
 toward zero). This singularity occurred at distances between $R_p$ and $m^{-1}$ depending on the mass term.
 In view of this singularity, we could not address the issue of ``shooting''
 the initial data with the aim of matching an exponentially decaying solution
 at infinity.

 We have also performed simulations in the
$\lambda$, $\mu$, $\nu$ gauge.  Their results have again all be negative: all
solutions starting from the center developed some strange behaviour for
large radii, and never showed any tendency to match the perturbative
solution at infinity.  Let us describe some of our results in the
$\lambda$, $\mu$, $\nu$ gauge.

 In view of the nature of the ODE system in this gauge we a priori need to give initial
 data, at $R=0$, for     $\lambda(0)$,
$\nu(0)$, $\mu(0)$ and $\mu'(0)$. First of all, we know that in order not to have a
conical singularity we should have $\lambda(0)=0$.
Second, if we remember that the function $\m$ defines the link between
the ``flat'' and the ``curved'' radial coordinates $ r = R e^{-\m(R)/2} $ , the regularity
(in Cartesian-like coordinates) of this link implies that we must also require
  $\mu'(0) = 0$.  Finally, we have only two initial data:  say $ C \equiv \nu(0)$ and  $D \equiv \mu(0)$.

 The leading terms for the three functions are:
\ba
&&\lambda=A R^2+\dots\\
&&\nu=C+BR^2+\dots\\
&&\mu=D+ER^2+\dots
\ea
In GR the quantities $A$ and $B$ are related to the density $\rho_0$ (for a constant
density star) and the central pressure $P_c$ as:
\ba
&&3A_{GR}=8 \pi G \rho_0\\
&&2B_{GR}-A_{GR}=8 \pi G P_c
\ea
while the constant $C$ is arbitrary. In the  massive gravity case, we obtain the following
relations:
\ba
&&3A=8 \pi G \rho_0+m^2 \Sigma_1(C,D)\label{sig1}\\
&&2B-A=8 \pi G P_c+m^2\Sigma_2(C,D)\label{sig2}
\ea
for all four potentials ${\cal V}^{(i)}$, $i=1,2,3,4$. The functions $\Sigma_1(C,D)$ and
$\Sigma_2(C,D)$ are zero when we sit on the point $C=D=0$, but they may have non-trivial
roots (we will see more in the next section). In order to study the conjecture
of Vainshtein (which assumes that we are near GR around the source)
 we explored the region of the  $(C,D)$
parameter space that is in the neighbourhood of $(0,0)$,
namely  $ -0.1 \leq C \leq 0.1 $ and $  -0.1 \leq D \leq 0.1$.

For such a choice of initial data,  one finds that, for small enough $m$, the solution
stays very close to the GR result until a radius which is very roughly of order $R_V$ and then the solution starts to deviate considerably.
 Note that these numerical simulations provide still another way of constructing {\it local} solutions
 of massive gravity that are close to GR ones. Note also that, when using a matter density model where
 $\rho$ undergoes a finite jump at $R = \R$, the variables $\l', \n''$ and $\m''$ suffer
 corresponding, correlated jumps.

However, even when changing in a continuous manner the initial data, we never found
any solution which, at large radii, tend to match the perturbative one.
In fact, all solutions tend to run into a numerically unstable behaviour, where the
simulation stops after a finite radius. We saw no tendency that the functions at distances much bigger than $R_p$
match the perturbative solution ones. We could not run the simulation far enough and
could not understand if this denotes that there is a singularity at some finite distance from the star. However, by numerically transforming the $\l$, $\m$, $\n$ variables into the 
$c$, $\bar{c}$, $\bar{b}$ ones we have found that at the point of numerical 
difficulties of the $\l$, $\m$, $\n$ gauge, there was a tendency for the $c$, 
$\bar{c}$, $\bar{b}$  variables to exhibit their usual singular behaviour 
($\bar{b} \to 0$, $\dot{c}~\dot{\bar{c}} \to 0$).

Summarizing:  Our numerical simulations have falsified the conjecture put forward in
 \cite{Vainshtein:sx} and  \cite{Deffayet:2001uk}.  Far from helping to regularize the
  discontinuities   and attendant     $\O(m^{-n})$  terms exhibited by the linearized
  approximation,  the non-linear effects in massive gravity aggravate the situation
 by  developing singularities at a finite radius when starting from the unique,
 asymptotically flat solution.

 This negative result closes, in our opinion, one possibility. However, it still leaves open
 several possibilities for eventually reconciling massive gravity with phenomenology.
 Indeed,  our study so far has focused on only one type of possible asymptotic
 behaviour at infinity: namely, \Ref{bc}, {\it i.e.} the case where the physical metric
 asymptotes the ``trivial'' vacuum solution. In the next section, we shall
 point out that even the simple-minded mass terms \Ref{V1}-\Ref{V4} might
 admit other  (translationally invariant) vacuum solutions. Finally, in the last
 section we shall consider the possibility that continuity be restored for
 solutions which asymptote some non-trivial cosmology at infinity.

\section{Non-trivial translationally-invariant, spherically symmetric vacua}

Returning to our original potentials, let us examine all their possible
translationally invariant, spherically symmetric vacua.  A translationally invariant
vacuum is a solution (in absence
of matter) for which, in some gauge, both $g_{\m\n}$ and   $f_{\m\n}$   are
constant ({\it i.e.} independent of the spacetime coordinates $x^{\m}$).
Requiring spherical symmetry  will impose a further constraint on the
symmetry group fixing the bi-metric configuration   ($g_{\m\n}$, $f_{\m\n}$).

The first constraint translates in the statement that the   matrix
$ g^{\m}_{\n} \equiv f^{\m\s} g_{\n\s}$ can be globally diagonalized as
 diag$(\l_0,\l_1,  \l_2 , \l_3)$, where the eigenvalues $\l_i$ are constant.
 In other words, there exist coordinate systems where the two metrics read:
 \be
 {\bf f} = - dt^2 + dx^2 + dy^2 + dz^2 \; , \;
  {\bf g} = - \l_0  dt^2 +\l_1 dx^2 +\l_2 dy^2 +\l_3 dz^2
\ee
Requiring spherical symmetry then imposes the constraint
$ \l_1= \l_2 = \l_3$. This leaves only two constants.

Finally, rewording the result in terms of the  $\lambda$, $\mu$, $\nu$ gauge,
one easily finds that the above requirements are equivalent to the
constraints:
\be
\label{vac0}
\l =0~~~,~~~  e^{\m} = \l_1 = const.~~~,~~~   e^{\n} = \l_0 = const.'
\ee
Technically, we are then left with looking for extrema of the mass terms with
constant values of   $\lambda$, $\mu$, $\nu$ of the type  \Ref{vac0}.

For the last three mass terms we found that they only admit the
trivial vacuum $ ( \l_1  , \l_0 ) = (    e^{\m} ,  e^{\n} ) = (1,1)$.

On the other hand, we found that the first mass term $ {\cal V}^{(1)} $ admits
three possible vacua. This can be seen by looking at the  $\Sigma_1(C,D)$, $\Sigma_2(C,D)$ functions in Eqs. \Ref{sig1}, \Ref{sig2}:
\ba
&&\Sigma_1(C,D)=-{3 \over 4}(e^D-1)(e^D+3e^C-2)\\
&&\Sigma_2(C,D)={1 \over 4}(5e^D(e^C-3)+7e^{2D}-3e^C+6)
\ea
remembering that $\n=C$, $\m=D$ in vacuum. Then the conditions $\Sigma_1(C,D)=0$, \mbox{$\Sigma_2(C,D)=0$} have three solutions:  the trivial vacuum   $ ( \l_1  , \l_0 ) = (    e^{\m} ,  e^{\n} ) = (1,1)$,
plus two new ones:
\be
 ( \l_1  , \l_0 ) = (    e^{\m} ,  e^{\n} )=  \left({1 \over 2}, {1 \over 2}\right) \;~~~ , ~~~ \;
 ( \l_1  , \l_0 ) = (    e^{\m} ,  e^{\n} ) = \left({3 \over 2}, {1 \over 6}\right)
\ee
Note that the first non-trivial vacuum corresponds to   $g_{\m\n} =  {1 \over 2} f_{\m\n}$.
Such a conformal situation respects not only spherical symmetry, but in fact also
the full Poincar\'e symmetry. By contrast, the last vacuum breaks the
 Poincar\'e symmetry to a mere Euclidean one.

Let us examine now the perturbations around these vacua for the first potential. For the 
trivial vacuum,  we already know that the linearized perturbations feature a Pauli-Fierz
mass term, {\it i.e.} a ratio $\a =1$ between the two terms, with  a spin-2 mass, say $m_2$,
identical to the mass parameter entering the action, {\it i.e.} $m_2^2=m^2$.
 For the  $ ({1 \over 2}, {1 \over 2})$ vacuum, one finds that   $\alpha=-{1 \over 8}$ and
$m_2^2=16 m^2$. The fact that we do not have a Pauli-Fierz mass term means,
as we recalled above, that we have a ghost-like scalar field ($h$). In addition, the
fact that the value of $\a$ is not between $1/4$ and $1$   means that this ghost degree of
freedom  is also tachyonic, since $m_0^2=-{2 \over 3}m_2^2$.
As we said above, such excitations are pathological, and we do not wish to consider
them (though we know that this is a cheap way of ensuring continuity as
$m^2 \to 0$).

 Finally, the third vacuum $({3 \over 2}, {1 \over 6})$, does not give an equation of the type
(\ref{BD}), but rather instead of the $h_{\mu \nu}-\alpha \eta_{\mu \nu} h$  mass term, 
one finds the following contribution:
\be
D^{\kappa \lambda}_{\mu \nu}h_{\kappa \lambda}+E_{\mu \nu} h \label{newmass}
\ee
where the tensors $D^{\kappa \lambda}_{\mu \nu}$ and $E_{\mu \nu}$ are constructed from the
two background metrics $\bar{\eta}_{\mu \nu}$ and $\eta_{\mu \nu}$, if $g_{\mu 
\nu}=\bar{\eta}_{\mu \nu}+h_{\mu \nu}$.

The important point is that these tensors for the background we are considering are {\it 
not} proportional to $\delta_{\kappa}^{\mu}\delta_{\lambda}^{\nu}+(\mu \leftrightarrow 
\nu)$ and $\bar{\eta}_{\mu \nu}$ respectively. This happens because the ``metric 
condensate'' is {\it not} proportional to the background of the observable metric, and therefore,
as we said, breaks the underlying Poincar\'e invariance. Thus, as we see from Eq. 
(\ref{newmass}), the equation of motion for the perturbation $h_{\mu \nu}$ will have 
different structure from Eq. (\ref{BD}). We leave to future work the task of
redoing, in this new setting, the analysis of    \cite{Boulware:my}.

In the next section, we shall finally study other types of vacua, which admit a cosmological interpretation.

\section{Continuous ``non-co-diagonal'' solutions with de Sitter asymptotics in
massive gravity}
\label{SSBsol}

In this section we explore another possibility for eventually reconciling massive
gravity with phenomenology. This possibility is based on an interesting
type of spherically symmetric stationary solutions of massive gravity found long ago by
 Salam and Strathdee  \cite{Salam:1976as}. [These types of solutions were then
 generalized to a full bigravity setting \cite{Isham:1977rj}.]
 These solutions tend, at spatial infinity, towards a de Sitter solution, instead of
 one of the translationally-invariant vacua considered above.

 The aim of this section is two-fold: (i) to describe the ``universality class''  \cite{Damour:2002ws}
 of mass terms which admit such solutions, and (ii) to show how the vacuum solution
 of    \cite{Salam:1976as} can be extended to a globally regular spacetime
 representing a ``star'' embedded in a de Sitter background.

 \subsection{Spontaneous symmetry breaking in massive gravity}

 Up to this section, we have only considered   spherically symmetric   solutions
 which could be represented in a co-diagonal way, {\it i.e.} such that both $\bf g$
 and ${\bf f}$ are diagonal in the same coordinate system. As we indicated,
 perturbation theory  (around the trivial vacuum)  necessarily select such
 a type of solution,  if the matter sources are supposed to be ``at rest''.
 However, non-perturbative  solutions  might introduce an interesting twist
 in the link between the two metrics       ${\bf g}$   and ${\bf f}$.
 A convenient formalism  for describing this possibility consists in introducing a ``link field'' $Y(x)$ (also called ``Stueckelberg field'') relating  the two metrics   ${\bf g}$   and ${\bf f}$. This formalism first appeared in \cite{Green:pa} and, in more detail in \cite{Siegel:1993sk}, and has been recently developed in \cite{Arkani-Hamed:2002sp}. More precisely,  
 instead of writing the theory of massive gravity in a coordinate system 
 which is common to both metrics, and of writing the mass term in 
  terms of the difference  $g_{\m\n}(x) - f_{\m\n}(x)$ between the two metric tensors
 (supposed to live on the same manifold, and expressed in the same coordinate system),
  one can explicitly introduce the map $Y$ between the
 ${\bf g}$-Riemannian manifold, and the   ${\bf f}$-one.  If we use independent coordinate
systems  on both  manifolds (say $x^{\m}$ on the  ${\bf g}$ manifold ${\cal M}_{ g}$, and  $X^{\a}$ on
the  ${\bf f}$-one ${\cal M}_{ f}$) the abstract map $Y$  takes the  explicit form
$X^{\a} = Y^{\a}(x^{\m})$.   This map (from ${\cal M}_{ g}$ towards ${\cal M}_{ f}$)  ``pulls back''  the
metric  ${\bf f}$ into an image metric $ Y_{*}{\bf f}$ which lives on ${\cal M}_{ g}$.
This leads to expressing the mass term in terms of:
\be
\label{HY}
H_{\m\n}[g,f,Y] =  g_{\m\n}(x) -    \de_{\m}Y^{\a}  \de_{\n}Y^{\b} f_{\a\b}(x)
\ee
Then the mass term (and therefore the action) becomes a functional of three
fields: $g, f$ and $Y$.  In the massive gravity context,  ${\bf f}$ is a non-dynamical
flat metric, and we can always choose to express it in Lorentzian 
coordinates, {\it i.e.} with     $ f_{\a\b}(x) = \e_{\a\b}$. Then the action depends only
on two dynamical fields:  $g$ and $Y$.  The usual formalism (used above) consists of
using the gauge invariance of the theory to fix $Y = $ id (which destroys the
explicit gauge invariance). Ref.   \cite{Arkani-Hamed:2002sp} advocates the usefulness
of not using this ``unitary gauge'', but instead of working with an explicitly gauge invariant
theory with two fields: $g$ and $Y$.  The immediate problem with this proposal is then that
the kinetic terms for the $Y$ field are very non-linear because they result from
replacing  \Ref{HY} in the mass terms. [For instance, the existence of a sixth
degree of freedom is hidden in the highly non-linear nature of this
kinetic term.]

 One can view $Y$ as a gravitational analogue of the Higgs field.
 Ref.   \cite{Arkani-Hamed:2002sp} has indicated how  to deal with
 perturbation theory  around the ``trivial  Higgs  configuration'' $Y_0^{\a}(x^{\m}) = x^{\a}$,
 by expanding the theory in powers of the ``Goldstone''  field $\p^{\a}(x)$ such that
 $Y^{\a}(x^{\m}) =   x^{\a}  + \p^{\a}(x)$.
 The analogue, in this language, of the perturbative co-diagonality of two SSS metrics is the
 following. If we consider some  static matter source (in  ${\cal M}_{ g}$),
 perturbation theory for the Goldstone field will never generate  a  time-component
  $\p^{0}(x)$ (which induces a non-zero mixed component   $g _{0i}(x)$ in the
  physical metric).

  In this language, one needs to be in a non-perturbative situation of ``spontaneous
  symmetry breaking'' to expect such a    $\p^{0}(x)$ to develop.
This situation was in fact recently considered, {\it mutatis mutandis}, in  \cite{Damour:2002wu}.
There it was noticed that besides the stable  family of bigravity cosmological solutions,
where the two metrics ${\bf g}_1$ , ${\bf g}_2$ can be written in a time-orthogonal gauge
($g_{0i \, 1} = 0 =g_{0i \, 2}$) and  where  the remaining metric coefficients ($g_{00}, g_{ij}$)
 depend only on time, there might also exist, under some conditions, families
 of ``symmetry breaking'' solutions where the relative ``shift vector'' $b^i$ (proportional to
$ g_2^{0i }$ in the gauge where  $ g_{0i \, 1} = 0$) is not equal to zero.  The possible
non-zero values of   $b^i$ were the extrema of the mass term
$V( {\bf b}) \sim V_0 + a  {\bf b}^2 + b  {\bf b}^4 + \O ( {\bf b}^6 )$. Similarly to the
Higgs mechanism,  for certain shapes of the function  $V( {\bf b})$ ({\it e.g.} when $a<0$
and $b>0$) there might exist, besides the trivial
extremum  ${\bf b} = 0$, non-trivial extrema  ($ {\bf b} \neq 0$) of   $V( {\bf b})$.
The precise condition for this to happen was studied in the Appendix B of  \cite{Damour:2002wu}.
We are going to recover the same condition for the possibility of a similar
symmetry breaking in the context of SSS solutions.

In this context, the co-diagonal form  \Ref{la1},   \Ref{la2} is not the most general
form compatible with spherical symmetry and stationarity.  The most
general form depends on 4 functions, and can be written in the
following        $\lambda$, $\mu$, $\nu$, $\a$ gauge:
 \ba
&&ds^2=-e^{\nu(R)}dT^2+e^{\lambda(R)}dR^2+R^2 d\Omega^2 \label{la1'}\\
&&ds_{\rm fl}^2=-( dT + \a dR)^2 +\left(1-{R\mu' \over 2}\right)^2e^{-\mu(R)}dR^2+e^{-\mu(R)}R^2
d\Omega^2\label{la2'}
\ea
Here all the functions $\lambda$, $\mu$, $\nu$, $\a$   depend only on $R$.
The above gauge has the advantage of separating the variables into two groups:
the ``physical'' variables       $\lambda$,   $\nu$  (which are directly observable in
gravitational experiments), and the ``gauge'' variables   $\mu$, $\a$ which enter only
the unobservable background flat metric. The gauge variables enter the action    \Ref{action}
{\it only through the mass term}     $ -(m^2 / 4) {\cal V}({\bf f}^{-1}{\bf g})$.
Therefore one can immediately obtain the equations of motion of    $\mu$, $\a$
by varying only the mass term:
\ba
&&{\d \over \d \m}   {\cal V}({\bf f}^{-1}{\bf g})  = 0  \label{meq'}\\
&&{\d \over \d \a}   {\cal V}({\bf f}^{-1}{\bf g})  = 0     \label{aeq'}
\ea
Equation \Ref{meq'} above generalizes the constraint \Ref{coneq} to the case where $\a \neq 0$. Let us consider here the new equation     \Ref{aeq'}.  A crucial point is that $\a$ enters ${\cal V}$
only algebraically.  Therefore, similarly to what we recalled above concerning the
 dynamics of the cosmological shift vector ${\bf b}$, non-trivial SSS solutions will exist
 only if    ${\cal V}(\a)$  admits non-trivial extrema with respect to $\a$.
 To understand what this implies for the ``universality class'' of such mass terms, let us
 relate the $\a$ dependence on the eigenvalues of the gravitational energy-momentum
 tensor   $T^{(g)}_{\mu\nu}$. Let us recall (from     \cite{Damour:2002ws,Damour:2002wu})
 the following  {\it bi-geometrical} facts: The bi-gravity configuration $({\bf g}, {\bf f})$
 defines, at each spacetime point, a preferred moving frame with respect to which
   ${\bf g}   = {\rm diag} (-\l_0,\l_1,\l_2,\l_3)$ and    ${\bf f}   = {\rm diag} (-1,1,1,1)$.
   [Note that this preferred frame is not proportional to the  coordinate frame used in the
   normal gauges  such as the   $\lambda$, $\mu$, $\nu$, $\a$ gauge.] The eigenvalues
   $\l_a$ of $ {\bf f}^{-1}  {\bf g}$  depend on $R$. They can be easily computed from
 the metric coefficients  $\lambda$, $\mu$, $\nu$, $\a$. For instance:
 \be
  \l_2= \l_3 = e^{\m(R)}
 \ee
 Let us focus on the $\a$ dependence of the eigenvalues. Note that   $\a$ disappears when
 calculating the determinant of    ${\bf f}$. Therefore the product:
 \be
 \D \equiv  \l_0 \l_1
 \ee
 which is the ratio of the two $2 \times 2$ determinants  of the $(T,R)$ sub-blocks
 of the two metrics, does not depend on $\a$.  Indeed,
 \be
 \D =  {e^{\n+\l+\m} \over            (1-R\mu' / 2 )^2 }
 \ee
On the other hand, $\a$ enters
 the trace of the   $(T,R)$ sub-block of   ${\bf g}^{-1}{\bf f}$. Calculating this trace
 yields:
 \be
 {1 \over   \l_0} +  {1 \over   \l_1} = e^{-\n} + e^{-\l} [ (1-{R\mu' \over 2} )^2  - \a^2] =
 {  \l_0    +  \l_1 \over   \l_0 \l_1}
 \ee
 When varying only $\a$, one therefore finds that   $ \l_2= \l_3$  and $\l_0 \l_1$
 do not change, but that   $\l_0 +  \l_1$ varies proportionally to $\a \d \a$.
 On the other hand, the variation of the mass term $ S^{(g)} \propto - {\cal V}$
 in the action is related, by definition, to the logarithmic changes
 of the eigenvalues  by $\d S^{(g)} = (1/2) \sqrt{g} { T^{(g)}}_{a}^{a} \d \log \l_a$,
 where    ${ T^{(g)}}_{a}^{a}$   denote the eigenvalues (in the preferred frame
 which is easily seen to diagonalize also  $T^{(g)}_{\mu\nu}$) of the gravitational
 energy tensor.  The final result is that the variational derivative  \Ref{aeq'}
 is found to be proportional to  the product:
 \be
 \a {    {T^{(g)}}_{0}^{0}   - { T^{(g)}}_{1}^{1}  \over  \l_0 - \l_1 }
\ee
As we said above, a universal solution of this $\a$-constraint is the trivial
case $\a =0$.  The spontaneous symmetry breaking (SSB) situation corresponds
to the case where it is the second factor above which vanishes.
Using the formulas given in    \cite{Damour:2002ws,Damour:2002wu} one can
easily see that the condition for  SSB  does not depend on the choice
of density  prefactor  in front of ${\cal V} \equiv w V$, where $w$ can be either
$\sqrt{-f}$, $ \sqrt{-g}$, $(fg)^{1/4}$ etc... Finally, we can write the condition for SSB
in terms of the scalar potential $V(\l_0,\l_1,\l_2,\l_3) \equiv  {\cal V} / w$  as  the existence of solutions
for the equation:
\be
\label{SSB}
\r (\l_0,\l_1,\l_2,\l_3) \equiv { 1 \over  \l_0 - \l_1} ({ \de V \over \de \ln \l_0 } - { \de V \over \de \ln \l_1 })  = 0
\ee
For instance, if we consider the class of potentials $V$ which depend only on $\s_1$ and $\s_2$,
where $\s_n \equiv  \S_a (\ln \l_a)^n $, we recover the condition $\de_{\s_2} V =0$ derived in
  \cite{Damour:2002ws,Damour:2002wu}.

 It happens that all the mass terms written  in \Ref{V1}-\Ref{V4} can satisfy the SSB condition.
 First, we note that they contain only two independent ``scalar'' potentials $V$. Moreover,
 the two types of scalar potentials can be mapped into each other by the exchange
 $ {\bf g} \leftrightarrow   {\bf f}  $, which is equivalent to the exchange $ \l_a  \leftrightarrow  1/\l_a$.
 Therefore, it suffices to consider the potential $V^{(1)}$ defined, say, by:
\be
V^{(1)} \equiv {1 \over 2}  \left\{{\rm tr}[(({\bf g}-{\bf f}){\bf f}^{-1})^2]-({\rm tr}[({\bf
g}-{\bf f}){\bf f}^{-1}])^2 \right\} = {1 \over 2}[ \sum_a (\l_a-1)^2 -  (\sum_a (\l_a-1) )^2]
\ee
The computation of the quantity $\r$ defined above then yields:
\be
\r = 1 - \sum_{c \neq 0,1} (\l_c -1) = d -  \sum_{c \neq 0,1} \l_c
\ee
where $d$ denotes the space dimension.  In our case, it is $d=3$, and we can also use
$\l_2 = \l_3$ to write $\r = 3 - 2 \l_2$.

Finally, we conclude that for the mass terms  $ {\cal V}^{(1)}$ and   $ {\cal V}^{(2)}$:
\be
\r[ {\cal V}^{(1)}] =  \r[ {\cal V}^{(2)}] =   3 - 2 \l_2
\ee
while the corresponding results for the other ones (obtained by exchanging    $ \l_a  \leftrightarrow  1/\l_a$)
read:
 \be
\r[ {\cal V}^{(3)}] =  \r[ {\cal V}^{(4)}] =  -3 + 2/ \l_2
\ee
Therefore the SSB condition ({\it i.e.} the equation of motion for $\a \neq 0$) can be satisfied
for all four potentials. In the first two cases, it fixes the value of $\l_2$ , {\it i.e.} of $e^{\m}$,
to be simply $3/2$, while in the last two cases one must have   $\l_2 = e^{\m} = 2/3$.
[The corresponding values in space dimension $d$ would be $d/(d-1)$ and $(d-1)/d$,
respectively.]
It is a happy technical accident (which simplifies the subsequent derivations) that these values are constant.
This is, however, not crucial. The essential point is that the SSB condition
$\r(\l_a) =0$ admits solutions.  More general types of SSB potentials might
admit solutions of    $\r(\l_0, \l_1,\l_2) =0$  along some curved  hypersurface  in
$\l_a$-space, say $\l_2 = f( \l_0, \l_1).$

\subsection{Non-codiagonal solutions in massive gravity}

Let us now study the global structure of SSB solutions (both in the exterior and the interior).
There are two ways of proceeding. Either we continue using a ``geometrical'' approach
based on the systematic use of the invariant eigenvalues, or we use a more pedestrian
approach based on writing down explicitly all the field equations in some gauge. Before using a more
pedestrian approach let us briefly indicate the results of the ``geometrical'' one
in the simple {\it exterior case}.

  Having obtained the value of $\l_2 = e^{\m}$ we can now get a
simple constraint on the product of eigenvalues $\D    \equiv  \l_0 \l_1$.
Indeed, we found above (as a consequence of the SSB condition) that the
$2 \times 2$ $(T,R)$ block of the gravitational energy tensor satisfies (in its
diagonalizing frame) ${T^{(g)}}_{0}^{0}   ={ T^{(g)}}_{1}^{1}$. This condition
actually means that, {\it in any frame}, the  $2 \times 2$ $(T,R)$ block of
 $T^{(g)}_{\mu\nu}$ is proportional the  $2 \times 2$ unit matrix. When going back
 to the physical gauge     $\lambda$, $\mu$, $\nu$, $\a$ this means that the
 equation of state of the gravitational energy satisfy $ P_r^{(g)}= - \r^{(g)} $.
 Using this information in the usual Einstein equations (in the exterior)
 \Ref{laeq}, \Ref{nueq} gives us the simple constraint that the product
 $e^{\n} e^{\l}$, {\it i.e.}, modulo a constant, the quantity  $\D    \equiv  \l_0 \l_1$,
 is {\it constant}.  At this stage we know all the eigenvalues modulo the knowledge
 of, say, $\l_1(R)$. To get the radial variation of  $\l_1(R)$ it suffices to write down
 the conservation equation   \Ref{conseqaniso} for the {\it gravitational}
 part of the energy-momentum tensor. This is simplified by the result    $ P_r^{(g)}= - \r^{(g)} $
 and now reads
 \be
 \de_R P^{(g)}_r+{2 \over R}(P^{(g)}_r -P^{(g)}_t) =0
 \ee
 As the components  $ (- P_r^{(g)},P_r^{(g)},P_t^{(g)},P_t^{(g)})$ of the gravitational energy-momentum tensor
 are checked to depend only on the eigenvalues, the above equation gives an evolution equation
 for   $\l_1(R)$. Actually, one can see immediately a particular solution: namely
  $\l_1 = \l_2$. [One then checks that this is the only possible solution.]
   Indeed,  the isotropy condition $ \l_1 = \l_2= \l_3$ guarantees that
$P^{(g)}_r = P^{(g)}_t $. Then we get a solution of the above equation because all
eigenvalues are constant (indeed, we know already that   $\l_2= \l_3$ is constant,
and that the product $\l_0\l_1$ is also constant).

Finally, we end up (in the exterior region) with 4 constant eigenvalues, three of which are
determined to be $\l_1= \l_2= \l_3 = 3/2$ , or $2/3$ (depending on the choice of mass term),
and the fourth $\l_0$ being an arbitrary constant  (with $ \l_0 \l_1 = \D = const.$).
Though the set of eigenvalues introduces a dissymmetry between the (eigen-)time direction
and the space directions, there is a further simplification which comes (again because
of the SSB constraint) from the result above   ${T^{(g)}}_{0}^{0}   ={ T^{(g)}}_{1}^{1}$.
This condition, together with the equality of the three ``spatial'' eigenvalues,
shows that, finally $ {T^{(g)}}_{0}^{0}   = {T^{(g)}}_{1}^{1} =   {T^{(g)}}_{2}^{2}   ={ T^{(g)}}_{3}^{3}$.
Therefore the gravitational energy tensor is fully isotropic in spacetime:
${T^{(g)}}_{\m}^{\n}  = - \Lambda  \d _{\m}^{\n} $, where the constant  $ \Lambda$
is some function of the eigenvalues. Such a gravitational energy tensor is equivalent to
a cosmological constant   $ \Lambda$.  Therefore we conclude, without calculations,
that the only SSB spherically symmetric stationary solutions of massive gravity  for
the potentials \Ref{V1}-\Ref{V4}  must be Schwarzschild-(Anti-)de Sitter solutions.
This was, indeed, the result of   \cite{Salam:1976as}.

Let us now  generalize this exterior solution to a global solution, including  the interior of a star.
For concreteness, we shall consider the specific mass term ${\cal V}^{(4)}$, \Ref{V4}.  The
 main message of
this section would not be altered if we used any of the (\ref{V1}), (\ref{V2}) or
(\ref{V3}) instead.  We know already, from the previous reasoning, the general structure
of the exterior solution. We only need to compute now the deviation from  Schwarzschild-de Sitter (SdS)
in the interior of the star.
We could do the explicit calculations of the interior structure  by working, as above, with
the geometrical eigenvalues, and the corresponding  gravitational energy tensor. We shall however use a more
pedestrian approach, writing down the equations of motion in the gauge defined by the line elements
\Ref{la1'}, \Ref{la2'}. [In Appendix B we also present the results following the notation of  the
original derivation   \cite{Salam:1976as}.] For simplicity we will consider a star of isotropic pressure $P$. The $(T,T)$ and $(R,R)$ equations of motion, after setting $e^{\mu}=2/3$,  then read:
\ba
&&e^{-\lambda}\left[{\lambda' \over R}+{1 \over R^2}(e^{\lambda}-1)\right]=8 \pi G \left({9m^2 \over 64 \pi G}\sqrt{3 \over 2}~e^{-3(\l+\n)/2} + \rho\right) \label{issla}\\
&&e^{-\lambda}\left[{\nu' \over R}+{1 \over R^2}(1-e^{\lambda})\right] =8 \pi G \left(-{9m^2 \over 64 \pi G}\sqrt{3 \over 2}~e^{-3(\l+\n)/2} + P\right) \label{issnu}
\ea
The  conservation of the matter energy-momentum tensor gives the familiar equation:
\be
P' = - {{\n}' \over 2} ( \rho+P) \label{issP}
\ee
while from  Eq. \Ref{meq'} (which replaces the previous constraint $f_g =0$ derived from
the conservation of the energy-momentum tensor generated by the mass term) one gets an {\it algebraic}
equation for $\a$ with solution:
\be
\a^2(R) ={1 \over 4}e^{-\n}\left[3R (\l'+\n')-2(e^{\l}-1)(3e^{\n}-2)\right]\label{a2sol}
\ee
Note that $\a^2(R)$ undergoes a finite jump (downwards) as $R$ crosses the radius
of the star (from the interior).

In the exterior, one trivially finds by adding Eqs. \Ref{issla} and \Ref{issnu} that $\D'={2 \over 3}(e^{\l+\n})'=0$ which
 implies, as we said earlier, that $\D= {2 \over 3}(e^{\l+\n}) = const. \equiv\D_0 >0$. Then from Eq. \Ref{issla} for
 example one finds the usual SdS solution:
\be
e^{-\l} = {2 \over 3 \D_0} e^{\n} =  1- {R_S \over R} - {\L \over 3}R^2   ~~~,~~~{\rm with} ~~~ \L={3 \over 4 }~{m^2 \over  \D_0^{3/2}}
\ee
Then from Eq. \Ref{a2sol} one can directly find $\alpha$:
\be
\a^2={2 \over 3 \D_0}(e^{\l}-1)\left(e^{\l} -{9\D_0 \over 4} \right)
\ee
The requirement of positivity of $\a^2$ then restricts $0< \D_0<4/9$. Note also that at the position of the dS
horizon, the gauge function $\a$ diverges since $e^{\l} \to \infty$. However, this is a mere coordinate singularity
of the bigravity configuration $ \bf f$, ${\bf g}$ as is expected from the fact that the eigenvalues of
${ \bf f}^{-1} {\bf g}$  are regular there, being equal, in the exterior, to
$(\l_0,\l_1,  \l_2 , \l_3) = (3 \D_0/2, 2/3,2/3,2/3)$. A coordinate transformation which regularizes
{\it both}   $ \bf f$ and  ${\bf g}$ is $dT \to dT + \a(R)dR$. [Note that the black-hole solution of  \cite{Salam:1976as} is also regular at the event horizon by the same transformation, something not noticed in \cite{Salam:1976as}.] 

The exterior massive gravity solution above involves, as separate scales, only
$R_S$ and $\D_0^{3/4} \l_m$. Note, however, that (when $\D_0 \sim 1$) the scale $R_p$ \Ref{pscale} enters
indirectly the solution as the scale where the two separate contributions to  $ e^{-\l} = {2 \over 3 \D_0} e^{\n}$
become comparable to each other.

In the interior of the star, one has to solve the closed system of  differential equations \Ref{issla}, \Ref{issnu} and \Ref{issP}. Note that the contribution of the massive graviton in the interior equations of motion differs from a pure cosmological constant.  A crucial feature of this system is that the
graviton mass $m^2$ appears only with a positive power, and in front of
a lower-derivative term. General theorems on the continuity (with respect to parameters) of solutions of ODE systems
then guarantees that, if $m^2 \R^2 $ is small enough, any regular
interior solution in the GR limit    will be smoothly deformed into some
massive gravity interior one (satisfying the regularity condition
$\l(R) \propto R^2$ at the center). The only requirement then that the solution is acceptable is that $\a^2$
in the interior be positive. We have numerically verified that this criterion is not difficult to satisfy.

Finally, let us note that the only effect of the mass of the graviton in this kind of SSB solutions is the introduction of a cosmological tail with an associated scale $\sim \D_0^{3/4}\l_m$. Since  $\D_0$ is an arbitrary dimensionless integration constant (lying in the interval $0<\D_0<4/9$), one obtains the peculiar result that the cosmological scale of the theory is  undetermined. A priori, however, one should expect that $\D_0 \sim \O(1)$ and that its precise  value  should finally be determined when trying to match this SSS solution with a cosmological one. Note also, that since the regular SSB solutions that we constructed differ from the GR ones in the presence of a cosmological constant $\L$ only inside the star, one does not have observable consequences of the type claimed in  \cite{Dvali:2002vf,Lue:2002sw} that might test the model experimentally.

\section{Conclusions}

We have explored the issue of continuity in the purely massive gravity theory for several
mass terms, which all reduce to a Pauli-Fierz one in the linearized limit.
 We have discussed in detail the claim of \cite{Vainshtein:sx} that  the resummation of
 non-linear effects make the massless limit smooth.  We showed that the
 type of  local expansion (considered in some bounded domain outside the source)
  proposed in    \cite{Vainshtein:sx} does not exist for
 the simplest, action-based, mass term considered in the subsequent,
 more detailed publication \cite{ Deffayet:2001uk }. However, it exists for other
 mass terms.
 We showed the double series nature
of the expansion that was implicitly performed in \cite{Vainshtein:sx} and noted that 
different mass terms give different kinds of expansions.

Then we discussed the   global nature of massive gravity solutions.
When considering the inward radial evolution of general asymptotically
flat solutions, we find that they all end up in a singularity at some
finite radius well outside the source.  When, on the other hand,
we consider the asymptotics of a two-parameter class of solutions which are
close to GR in and near the source, we also find that they end up
in a singularity at some large radius. These results were obtained in two
different gauges, one of which allows one to qualitatively relate the
singular runaway of the solution    to a similar runaway found in recent
cosmological studies of massive gravity   \cite{Damour:2002wu}.

These results falsify the continuity claim (as it was formulated) of   \cite{Vainshtein:sx, Deffayet:2001uk},
and vindicate the statements of  \cite{vanDam:1970vg,Zakharov,Boulware:my} that
all asymptotically flat solutions of massive gravity are discontinuous as $m \to 0$,
and phenomenologically excluded.  Let us note, for completeness, that to fully
vindicate this negative result one should still explore the properties of the
new types of (asymptotically flat) vacua that  we have discovered in section 8 above.

In addition,  one should also explore the effect of using ``more confining'' potentials,
such as the potential $ V \propto  \s_2 - {\s_1}^2  + \l  {\s_2}^2  $ considered in   \cite{Damour:2002wu}.
There it was shown that such a mass potential can cure the runaway of the gauge
variable $e^{\g}$ (when $\l >0$).  One could hope that, within our context, it might
similarly cure the $\bar{b}$ runaway by confining   $\bar{b}$ within a finite
interval separated from zero.  However, we do not think that such a modification
of the mass term is sufficient for curing the discontinuity problem.  First, we anticipate that any
modification of the mass term by additional ``confining'' terms containing only parameters of order unity
(such as     $ \l$ in the example above)  will have great difficulty in confining   $\bar{b}$
because it will have to fight against the extremely large
dimensionless number $ h \sim G T/ m^2 \sim U/ (mR)^2$ which, according to perturbation theory,
will tend to enter the evolution of the gauge variables $ \bar{b} \sim \m \sim h $.
[And, indeed, our preliminary numerical simulations found a runaway   of   $\bar{b}$
toward zero with potentials  $ V \propto  \s_2 - {\s_1}^2  + \l  {\s_2}^2  $.]
Second,   even if one ``confines by brute force'' the evolution of  $\bar{b}$
by adding sufficiently efficient terms in $V$,  we see no reason why the resulting global
solution should get close to a GR one.  We anticipate that, even if the runaway singularity
is cured, the resulting (non spontaneously symmetry breaking) solutions will
be  very different from GR solutions, and therefore phenomenologically unacceptable.

Our negative result still leaves open the possibility, proposed in  \cite{Damour:2002ws, Damour:2002wu},
that continuity and phenomenological compatibility be restored by considering solutions
that  asymptote some non-trivial cosmological spacetime of curvature $ {\cal R} = {\O}(m^2)$.
[Note that this proposal differs from what is suggested by the continuity results
in (A)dS backgrounds  \cite{Higuchi:1986py,Higuchi:gz,Kogan:2000uy,Porrati:2000cp}
which need $ {\cal R} \gg{ \O}(m^2)$ to restore continuity.]
We have provided evidence for this possibility by exhibiting a particular class of
``symmetry breaking'' solutions with matter sources, which generalize black-hole-type solutions
found long ago in   \cite{Salam:1976as}.  As $ m \to 0$, these solutions smoothly tend
to an asymptotically flat  general relativistic spherical star  model. The ``price'' for continuity is, however, twofold:
(i) asymptotically the solution tends, when $m \neq 0$,
 to a de Sitter solution, and (ii) the mass term must belong to one
of the special ``universality classes'' delineated in   \cite{Damour:2002ws}.
The ``price'' (i) might actually be seen to be a virtue, as current cosmological data
favor such an asymptotic state for our universe. On the other hand, the
 requirement (ii) is not innocent because it was shown in the latter reference
that the mass terms coming from brane models do not belong to the needed
``symmetry breaking'' universality class.

Let us finally note that our positive results (on the re-establishment of continuity
for symmetry-breaking  asymptotically dS solutions) are still far from a satisfactory
proof of the physical admissibility of massive gravity theories.  Many subtle issues
need to be checked, such as: (i) can the simple spherically symmetric solutions
explored here be deformed into  inhomogeneous spacetimes able to describe the
solar system and the observed cosmological  universe ?
(ii) are such solutions  attractors   of the cosmological
evolution ?, (iii)
are they stable under  quantum fluctuations\footnote{See \cite{Deser:2001xr} for recent studies
of massive spin fields in (A)dS backgrounds.}  ?  In particular, we worry about
the (classical and quantum) effect of the ``sixth degree of freedom'' present in
massive gravity  \cite{Boulware:my} whose r\^ole in the simple solutions explored so far
might have been unduly downplayed.

 Let us finally mention that a better setting
for investigating these questions (especially the ``cosmological attractor'' issue)
would be a full {\it bi-gravity} theory      \cite{Isham:gm,Damour:2002ws},
which, contrary to the purely massive gravity explored here,
 does not contain any{\it a priori} fixed, non-dynamical background. In that case, there exist
  bi-de Sitter ``locked''
 solutions \cite{Damour:2002wu} which are cosmological attractors (and which are
 more general than the above considered SSB ones). They can be trivially generalized to
 bi-Schwarzschild-deSitter solutions with correlated source masses. An interesting question then is whether these
  kinds of solutions are  ``spacelike'' attractors as well, if we detune the special relation between the source masses.

\vskip1cm

\textbf{Acknowledgments:}
We would like to thank B. Tekin and W. Siegel for bringing to our attention Refs. \cite{Jun:hg} and \cite{Green:pa,Siegel:1993sk}.
I.K. is supported in part by PPARC rolling grant PPA/G/O/1998/00567 and the EC TMR grants HPRN-CT-2000-00152 and  HRRN-CT-2000-00148.
A.P. is supported in part by the European Community's Human Potential
Programme under contracts HPRN--CT--2000--00131 Quantum Spacetime,
HPRN--CT--2000--00148 Physics Across the Present Energy Frontier
and HPRN--CT--2000--00152 Supersymmetry and the Early Universe.
I.K. and A.P. wish to thank IHES for hospitality during the inception of this work.
T.D.  dedicates this work to the memory of F. Damour.

\vskip1cm

\def\theequation{A.\arabic{equation}}
\setcounter{equation}{0}
\vskip0.8cm
\noindent
{\Large \bf Appendix A: The functions $f_t$, $f_R$ and $f_g$}
\vskip0.4cm
\noindent

In this Appendix we will present the functions $f_t$, $f_R$ and $f_g$ for the various 
massive gravity theories that we considered in the text. First, these quantities for the 
postulated energy-momentum $T^{(V)}_{\mu \nu}$ (see (\ref{VTmn})) of \cite{Vainshtein:sx} 
read:
\ba
&&f_t=-{1 \over 2}\left[e^{\mu+\lambda} \left(1-{R \mu' \over 2}\right)^{-2}+2 
e^{\mu}-3\right]\\
&&f_R=-{1 \over 2}e^{-\mu}\left(1-{R \mu' \over 2}\right)^{2}(3-2e^{\mu}-e^{\nu})
\ea
\ba
f_g&=&{e^{-\lambda-\mu}\over 2 R}\left[1-e^{\lambda}\left(1-{R \mu' \over 
2}\right)^{-2}\right]\left[e^{\lambda+\mu}+(e^{\nu}+2e^{\mu}-3)\left(1-{R \mu' \over 
2}\right)^2\right]\nonumber\\
&&+{1 \over 4}\nu' e^{-\nu}\left[e^{-\lambda-\mu+\nu}(e^{\nu}+2e^{\mu}-3)\left(1-{R \mu' 
\over 2}\right)^2-e^{\lambda+\mu}\left(1-{R \mu' \over 
2}\right)^{-2}-2e^{\mu}+3\right]\nonumber\\
&&+{1 \over 4}e^{-\lambda-\mu}\left[R \mu'^2 (e^{\nu}-3)-2\lambda'\left(1-{R \mu' \over 
2}\right)(e^{\nu}+2e^{\mu}-3)\right.\nonumber\\
&&\phantom{--------}-\mu'\left(4(e^{\mu}-3)+e^{\nu}(4+R\nu')\right)\nonumber\\
&&\left.\phantom{+++++++++1 
\over3}+6R\mu''-4e^{\mu}R\mu''+2e^{\nu}(\nu'-R\mu'')\right]\left(1-{R \mu' \over 2}\right)
\ea

For the theories which we have been studying, and for which the equations of motion stem 
from an action principle we obtain the following expressions. 

For the potential ${\cal V}^{(1)}$ (see (\ref{V1})) we have:
\ba
&&f_t=-{1 \over 4}e^{\nu}\left[e^{\lambda+\mu}(3e^{\nu}+2e^{\mu}-3)\left(1-{R \mu' \over 
2}\right)^{-2}+3e^{\nu}(2e^{\mu}-3)+e^{\mu}(e^{\mu}-6)+6\right]~~~~\\
&&f_R={1 \over 4}e^{\lambda}\left[3e^{\lambda+\mu}(2e^{\mu}+e^{\nu}-3)\left(1-{R \mu' 
\over 2}\right)^{-2}+e^{\nu}(2e^{\mu}-3)+e^{\mu}(e^{\mu}-6)+6\right]
\ea
\ba
f_g&=&-{e^{\mu} \over 8R}\left[-8(e^{\lambda}-1)(e^{\mu}+e^{\nu}-3)+3R^2\mu'^2(4e^{\lambda+\mu}+e^{\nu}(e^{\lambda}+4)+4e^{\mu}-3e^{\lambda}-12)\phantom{2 \over 3}\right.\nonumber\\
&&~~~~~~~~~+{1 \over 2}R^3 
\mu'^3(e^{\mu}+e^{\nu}-3)(R\mu'-8)-6R\lambda'e^{\lambda}(2e^{\mu}+e^{\nu}-3)\left(1-{R 
\mu' \over 2}\right)\nonumber\\
&&~~~~~~~~~+R^2 \mu'\nu'e^{\lambda}(2e^{\mu}+3e^{\nu}-3)+8R 
\mu'(6+3e^{\lambda}-2e^{\mu}-e^{\nu}(e^{\lambda}+2)-4e^{\lambda+\mu})\nonumber\\
&&~~~~~~~~~\left.\phantom{1 \over 2}-6R 
e^{\lambda+\nu}(\nu'+R\mu'')-2Re^{\lambda}(2e^{\mu}-3)(\nu'+3R\mu'')\right]\left(1-{R \mu' 
\over 2}\right)^{-3}
\ea

For the potential ${\cal V}^{(2)}$ (see (\ref{V2})) we have:
\ba
&&f_t={1 \over 2}e^{(3\nu-3\mu-\lambda)/2}\left[e^{\lambda+\mu}\left(1-{R \mu' \over 
2}\right)^{-1}+(2e^{\mu}-3)\left(1-{R \mu' \over 2}\right)\right]\\
&&f_R=-{1 \over 2}e^{(3\lambda-\mu-\nu)/2}(2e^{\mu}+e^{\nu}-3)\left(1-{R \mu' \over 
2}\right)^{-1}
\ea
\ba
f_g&=&{e^{-(\lambda+3\mu+\nu)/2} \over 
8}\left[-2\lambda'e^{\lambda+\mu}(2e^{\mu}+e^{\nu}-3)\left(1-{R \mu' \over 
2}\right)+{2e^{\nu} \over R}(4e^{\mu}+R\nu'(2e^{\mu}-3))\left(1-{R \mu' \over 
2}\right)^3\right.\nonumber\\
&&~~~~~~~~~~~~~~~~~~+{e^{\mu+\lambda} \over R}(-8(e^{\mu}-3)-4R 
\mu'(e^{\mu}+3)+R^2\mu'^2(2e^{\mu}+3)-2R^2\mu'(2e^{\mu}-3))\nonumber\\
&&~~~~~~~~~~~~~~~~~~-{e^{\lambda+\mu+\nu} \over 
R}(8+2R\nu'+R\mu'(R\mu'-R\nu'-4)+2R^2\mu'')\nonumber\\
&&~~~~~~~~~~~~~~~~~~~~~~~~~~~~~~~~~~~~~~~~~~~~~~~~~~~~~\left.+{8e^{\mu} \over 
R}(e^{\mu}-3)\left(1-{R \mu' \over 2}\right)^3\right]\left(1-{R \mu' \over 2}\right)^{-2}
\ea

For the potential ${\cal V}^{(3)}$ (see (\ref{V3})) we have:
\ba
&&f_t={1 \over 4}e^{-2\mu}\left[e^{-\lambda}(e^{\mu}+e^{\nu}(3e^{\mu}-2))\left(1-{R \mu' 
\over 2}\right)^{2}+2e^{\mu}(3e^{\nu}+1)-3e^{2\mu}(2e^{\nu}+1)-e^{\nu}\right]~~~~~~~~~\\
&&f_R={1 \over 4}e^{-\nu -2\mu+\lambda}\left[e^{\n}(1+6e^{\mu}(e^{\mu}-1))-e^{\mu}(3e^{\mu}-2)+e^{-\lambda}(e^{\nu}(3e^{\mu}-2)-e^{\mu})\left(1-{R \mu' 
\over 2}\right)^{2}\right]~~~~~~~~~
\ea
\ba
f_g &=& -{e^{-\l-\n-2\m} \over 8R}\left\{e^{\m}\left[-8(e^{\l}-1)-2\left(1-{R \mu' \over 2}\right)R\l'-2R\n'+R\m'(R\n'+R\m'-8)-2R^2\m''\right]\right.\nonumber\\
&&~~~~~~~~~~~~~~~+e^{\n}\left[8(e^{\l}-1)(3e^{\m}-1)-R^2\m'^2(3e^{\m}-4)+2R\l'(3e^{\m}-2)\left(1-{R \mu' \over 2}\right)\right.\nonumber\\
&&\left.\left.\phantom{1 \over 2}~~~~~~~~~~~~~~~~~~~~~~~~+R\m'(3e^{\m}-2)(8+R\n')-2R(3e^{\m}-2)(\n'-R\m'')\right]\right\}\left(1-{R \mu' \over 2}\right)~~~~~~~~~~
\ea

For the potential ${\cal V}^{(4)}$ (see (\ref{V4})) we have:
\ba
&&f_t=-{1 \over 2}e^{-(3\lambda +5\mu+\nu)/2}\left(1-{R \mu' \over 
2}\right)\left[e^{\lambda}(3e^{\mu}-2)-\left(1-{R \mu' \over 2}\right)^2\right]\\
&&f_R={1 \over 2}e^{-(\lambda +7\mu+3\nu)/2}\left(1-{R \mu' \over 
2}\right)^3\left[e^{\nu}(3e^{\mu}-2)-e^{\mu}\right]
\ea
\ba
f_g &=& -{e^{-(3\n+3\l+7\m)/2} \over 16 R}\left\{e^{\m}\left[2\left(1-{R \mu' \over 2}\right)\left(8-6R\l'\left(1-{R \mu' \over 2}\right)-6R\n'-6R^2\m''\right.\right.\right.\nonumber\\
&&\left.\left.\phantom{1 \over 2}~~~~~~~~~~~~~~~~~~~~~~~~~~~~~~~+R\m'(3R\n'+5R\m'-20)\right)+4e^{\l}(R\n'(3e^{\m}-2)-4)\right]\nonumber\\
&&~~~~~~~~~~~~~~~~~~~~~+e^{\n}\left[-2\left(1-{R \mu' \over 2}\right)\left(-8+24e^{\m}-R\m'(60e^{\m}-44)\phantom{1 \over 2}\right.\right.\nonumber\\
&&\left.\left.\left.~~~~~~~~~~~~~~~~~+R^2\m'^2(15e^{\m}-14)-6R\l'(3e^{\m}-2)\left(1-{R \mu' \over 2}\right)\right.\right.\right.\nonumber\\
&&\left.\left.\left.\phantom{1 \over 2}~~~~~~~~~~~~~~~~~~~~~~~~~~~~~~~~~-6R^2\m''(3e^{\m}-2)\right)+16e^{\l}(3e^{\m}-1)\right]\right\}\left(1-{R \mu' \over 2}\right)~~~~~~~~
\ea

For the potential ${\cal V}^{(\sigma)}$ (see (\ref{Vs})) we have:
\ba
&&f_t=-{1 \over 
8}e^{(3\nu-3\mu-\lambda)/4}\left[3\mu(\mu+\nu+4)+\lambda(3\mu+\nu+4)\phantom{0\over 
0}\right.\nonumber\\&&\left.~~~~~~~~~~~~~~~~~~~~~~~~~~~~~~~-2(2\mu+\nu+4)\log\left(1-{R 
\mu' \over 2}\right)\right]\sqrt{1-{R \mu' \over 2}}\\
&&f_R={1 \over 
8}e^{(3\lambda-3\mu-\nu)/4}\left[\nu(3\mu+\lambda+4)+\mu(3\mu+3\lambda+8)\phantom{0\over 
0}\right.\nonumber\\&&\left.~~~~~~~~~~~~~~~~~~~~~~~~~~~~~~~-2(2\mu+\nu)\log\left(1-{R \mu' 
\over 2}\right)\right]\sqrt{1-{R \mu' \over 2}}
\ea
\ba
f_g&=& -{e^{-(\l+\n+3\m)/4} \over 32 \sqrt{1-{R \mu' \over 2}}}\left\{(\m'+R\m'')\left[\m\left(8+3(\l+\m)-4\log\left(1-{R \mu' \over 2}\right)\right)-4(\n+2\m)\right.\right.\nonumber\\
&&~~~~~~~~~~~~~~~~~~~~~~~~~~~~~~~~~~~~~~~~~~~~~~~~~~~~\left.+\n\left(4+\l+3\m-2\log\left(1-{R \mu' \over 2}\right)\right)\right]\nonumber\\
&&~~~~~~~~~~~~~~~~~~~~~+\left(1-{R \mu' \over 2}\right)\left[(3\m'+\n'+\l')\left(\m\left(8+3(\l+\m)-4\log\left(1-{R \mu' \over 2}\right)\right)\right.\right.\nonumber\\
&&~~~~~~~~~~~~~~~~~~~~~~~~~~~~~~~~~~~~~~~~~~~~~~~~~~~~~~~~~~~~~~~~~~~~~~\left.+\n\left(4+\l+3\m-2\log\left(1-{R \mu' \over 2}\right)\right)\right)\nonumber\\
&&~~~~~~~~~~~~~~~~~~~~~~~~~~~~~~~~~~~~~~~~+\left(8\n'+{32 \over R}\right)\left(\l-2\log\left(1-{R \mu' \over 2}\right)\right)\nonumber\\
&&~~~~~~~~~~~~~~~~~~~~~~~~~~~~~~~~~~~~~~~~-4(3\m'(\m+\n)+2\n'(\n-\m)+\l'(\n+3\m))\nonumber\\
&&~~~~~~~~~~~~~~~~~~~~~~~~~~~~~~~~~~~~~~~~-4\m'\left(8+3(\l+\m)-4\log\left(1-{R \mu' \over 2}\right)\right)\nonumber\\
&&~~~~~~~~~~~~~~~~~~~~~~~~~~~~~~~~~~~~~~~~\left.\left.-4\n'\left(4+\l+3\m-2\log\left(1-{R \mu' \over 2}\right)\right)\right]\right\}
\ea

\def\theequation{B.\arabic{equation}}
\setcounter{equation}{0}
\vskip0.8cm
\noindent
{\Large \bf Appendix B: The SSB solution in a different gauge}
\vskip0.4cm
\noindent

In this Appendix we present the SSB solution of section \ref{SSBsol} in the notation of  the  original derivation   \cite{Salam:1976as}. We now use the following gauge for the two metrics:
The background flat metric is written as:
\be
ds_{\rm fl}^2=-dt^2 +dr^2+r^2 d\Omega^2
\ee
while the observable metric is written in the following form:
\be 
ds^2=-C(r)dt^2+A(r)dr^2+2D(r)dtdr+B(r) d\Omega^2
\ee
The off diagonal term $D/C$  is equivalent (after going from the `flat'' time $dt$
to the ``curved'' one $ dT = dt - (D/C) dr $, and
changing $dr = (dr/dR) dR$ ) to the (gauge) variable $\a$ above.

Note the values of the eigenvalues of ${\bf f}^{-1} {\bf g}$ in this new notation:
\ba
  &&\D \equiv  \l_0 \l_1 = A C + D^2,  \label{D} \\
  &&\l_0 +  \l_1  = A + C,  \label{S} \\
  &&\l_2 = \l_3 = B/r^2
 \ea
For simplicity, we will consider a star with isotropic pressure $P$. The energy momentum tensor is
then:
\be
T_{\mu\nu}^{\rm (matt)}=(\rho+P)u_{\mu}u_{\nu}+P g_{\mu \nu}
\ee
where $u_{\mu}=(-\sqrt{C},{D \over \sqrt{C}},0,0)$ is the unit velocity along the timelike
static Killing vector $\xi_{\mu}=(-C,D,0,0)$. Explicitly, it reads:
\be
T^{\rm (matt)}_{\mu \nu}=\left(\begin{array}{cccc}\rho C&-\rho D&~&~\\-\rho
D&~(\rho+P){\Delta \over C}-\rho A~&~&~\\~&~&P B&~\\~&~&~&P B \sin^2 \theta
\end{array}\right)
\ee
The first step for finding the solution is to impose the SSB condition \Ref{SSB}
(valid when $\a \neq 0$, {\it i.e.} $ D \neq 0$). For the mass term   ${\cal V}^{(4)}$, \Ref{V4}
that we consider for concreteness, this yields:
\be
 \l_2 = \l_3 ={  B  \over r^2}  ={2 \over 3}
 \ee
Then we can follow the steps  indicated above to derive the exterior solution.
The exterior constancy of $\D$ came from considering the effect of  the
source combination $\rho^{(g)} +P_r^{(g)}$.  In the present notation this corresponds to
 using $A T^{(g4)}_{tt}+C T^{(g4)}_{rr}=0$, to get $A T^{\rm
(matt)}_{tt}+C T^{\rm (matt)}_{rr}=\Delta (\rho+P)$, which implies that $A R_{tt}+C R_{rr}=8
\pi G \Delta (\rho+P)$. The latter equation explicitly reads:
\be
\Delta'=8 \pi G{\Delta^2 \over C}r(\rho+P)\label{mattD'}
\ee
The conservation of the gravitational energy-momentum tensor $T^{(g4)}_{\mu \nu}$
then gives:
\be
3r \Delta'-9 \Delta^2+(6(A+C)-4)\Delta=0\label{gravcos}
\ee
while the conservation of matter energy-momentum tensor $T^{\rm (matt)}_{\mu \nu}$ gives:
\be
P'=-{1 \over 2} {C' \over C}(\rho+P)\label{mattcos}
\ee
Combining Eqs. (\ref{mattD'}) and (\ref{gravcos}) we obtain the following expression
for the combination \Ref{S}:
\be
A+C={3 \over 2}\Delta+{2 \over 3}-4 \pi G{\Delta \over C} r^2 (\rho+P)\label{A+C}
\ee
One finally needs  only one non-trivial field equation to close the system, {\it
e.g.} the (t,t) component of the Einstein equations:
\be 
\Delta^2(6-32 \pi G r^2 
\rho)-4\Delta(C+rC')+4rC\Delta'=3m^2r^2\sqrt{\Delta}\label{nontrivial}
\ee

The system of radial ODE's \Ref{mattD'}, \Ref{mattcos}, \Ref{nontrivial} gives,
for any given equation of state $ P = f(\r)$, a closed evolution system for the
variables $\D$, $P$ and $C$. A crucial feature of this system is that the
graviton mass $m^2$ appears only with a positive power, and in front of
a lower-derivative term (see    \Ref{nontrivial}).
General theorems on the continuity (with respect to parameters) of solutions of ODE systems
then guarantees that, if $m^2 \R^2 $ is small enough, any regular
interior solution in the GR limit    will be smoothly deformed into some  regular
massive gravity interior one.

 To complete this Appendix, let us give the explicit form, in the $A,B,C,D$ gauge of the
 exterior solution. It is given by $ \D = const. \equiv \D_0$,   $ A+C={3 \over 2}\Delta_0+{2 \over 3}$,
 and by the nontrivial equation for $C$ whose general solution is easily found to be:
 \be
C={3 \over 2}\Delta_0 (1-p(r))
\ee
where
\be
p(r)={\tilde{R}_S \over r}+{m^2 \over 6 \Delta_0^{3/2}}r^2
\ee
 One then recognizes the    Schwarzschild-de Sitter (rather than AdS) solution  by changing the
 time variable to:
 \be
d\tilde{T}={1 \over \sqrt{\b+1}}\left(dt-dr {\sqrt{p(p+\b)} \over 1-p}\right)
\ee
 where $\b=(4 / 9\Delta_0)-1$. The observable metric gets transformed to an exact
Schwarzschild-de Sitter metric (here represented in a  conformally scaled form):
\be
ds^2={2 \over 3}\left[-(1-p)d\tilde{T}^2+{1 \over 1-p}dr^2+r^2 d\Omega^2\right]
\ee
The connection of these coordinates with the ones used in section \ref{SSBsol} is simply $R^2={2 \over 3}r^2$
and $dT={4 \over 9 \D_0}d\tilde{T}$. Note also that $\tilde{R}_S=\sqrt{3 \over 2}R_S$.

\end{document}